\documentclass[structabstract]{aa} 

\usepackage{graphicx}

\usepackage{txfonts}
\usepackage{natbib}

\def\kms{\rm {km\,s$^{-1}$}}
\def\x{$\times$}
\def\etal{et~al.}
\def\cmsq{\rm {cm$^{-2}$}}
\def\cmcub{\rm {cm$^{-3}$}}

\def\h2o{H$_2$O}
\def\so2{SO$_2$}
\def\13co{$^{13}$CO}
\def\nh3{NH$_{3}$}
\def\hco+{HCO$^{+}$}
\def\water18{H$_2^{18}$O}
\def\deltav{$\Delta \upsilon$}
\def\trans{1$_{1,0}$\,--\,1$_{0,1}$}

\begin{document}    

\titlerunning {Water and ammonia abundances in S140 with the Odin satellite} 
   \title{Water and ammonia abundances in S140 with the Odin\thanks{Odin is a Swedish-led satellite
   project
   funded jointly by the Swedish National Space Board (SNSB), the Canadian Space Agency (CSA), the National Technology Agency
   of
   Finland (Tekes) and Centre National d'Etudes Spatiales (CNES). The Swedish Space Corporation was the prime contractor and
   also is
   responsible for the satellite operation.} satellite}
\authorrunning{C.M.~Persson \etal} 
   \author{C.~M.~Persson
          \inst{1},
M.~Olberg\inst{1}, 
\AA.~Hjalmarson\inst{1},
M. Spaans\inst{2}, 
J.H.~Black\inst{1}, 
 U.~Frisk\inst{3},
T.~Liljestr\"om\inst{4},
A.O.H.~Olofsson\inst{1, 5},
D.R.~Poelman\inst{6},
 and   Aa.~Sandqvist\inst{7}
}

   \offprints{carina.persson@chalmers.se}
   \institute{Onsala Space Observatory, Chalmers University of Technology, SE-439 92 Onsala, Sweden\\
              \email{carina.persson@chalmers.se}
 \and Kapteyn Astronomical Institute, Rijksuniversiteit Groningen, P.O. Box 800, 9700 AV Groningen, The Netherlands 
\and Swedish Space Corporation, PO Box 4207, SE-171 04 Solna, Sweden
\and Mets\"ahovi Radio Observatory, Helsinki University of Technology, Otakaari 5A, 02150 Espoo, Finland
\and GEPI, Observatoire de Paris, CNRS; 5 Place Jules Janssen, 92195 Meudon, France
\and SUPA, School of Physics and Astronomy, University of St Andrews, North Haugh, St Andrews KY16 9SS, U.K.
\and Stockholm Observatory, AlbaNova University Center, SE-10691 Stockholm, Sweden
}

   \date{Received September 7, 2008; accepted November 14, 2008}

  \abstract
   {}
{We investigate the effect of  the   physical environment  on water and ammonia abundances   across the S140  photodissociation  region (PDR)   with an embedded outflow.
}
{We have used the Odin satellite to 
obtain 
strip maps
of the ground-state rotational transitions of \emph{ortho}-water and \emph{ortho}-ammonia,   as well as CO(5\,--\,4) and  \13co(5\,--\,4) 
across the    PDR, and  \water18 in the central position.
A physi-chemical
inhomogeneous PDR model was used to  compute the   temperature and  abundance distributions for water, ammonia and CO.
A multi-zone escape probability method then calculated the level populations and intensity distributions. 
These results are  compared to a homogeneous model computed with an enhanced version of the {{\tt RADEX}} code.
} 
{\h2o, \nh3 and \13co show emission from an extended PDR with a narrow line width of $\sim$3\,\kms. 
Like CO, the 
water line profile is   dominated by
outflow emission, however,   mainly in the red wing.
Even 
though CO shows strong self-absorption,
no signs of self-absorption  are seen in the water line.  
\water18 is not detected.
The PDR model  suggests that
the water emission  mainly arises from 
the surfaces of optically thick, high density clumps with $n(\mathrm{H_2}$)$\ga$10$^{6}$\,\cmcub~and a clump water abundance, with respect to H$_2$, of 5\x10$^{-8}$.
The mean water abundance 
in the PDR is    5\x10$^{-9}$, and between $\sim$2\x10$^{-8}$\,--\,2\x10$^{-7}$ in the outflow derived from a simple two-level approximation.
The   {{\tt RADEX}} model    points to  a somewhat higher   average PDR water abundance  of  1\x10$^{-8}$.
At low temperatures deep in the cloud the water emission is weaker,  likely due to adsorption onto   dust grains, while  ammonia is still abundant.
Ammonia is also observed in  the extended clumpy PDR, likely   from   the same  high density and warm  clumps as water. The average ammonia abundance is   about the same as for water: 4\x10$^{-9}$ and  8\x10$^{-9}$ given by  the PDR model and {\tt RADEX}, respectively. The differences between the models most likely arise due to uncertainties in density, beam-filling and volume filling of clumps.
The similarity of water and ammonia PDR emission is also seen in the almost identical line profiles observed close to the bright rim. 
Around the central position,  ammonia also shows some outflow emission
although weaker  than   water  in the red wing.
Predictions of the \h2o \trans~and 1$_{1,\,1}$\,--\,0$_{\,0,\,0}$ antenna temperatures  across the PDR 
 are estimated
with our PDR model  
 for 
the forthcoming observations with
the
Herschel Space Observatory.
}
   {}


   \maketitle

\section {Introduction}

The chemistry and the evolution of the interstellar medium  
depend to a high degree on the available amount of carbon and oxygen. 
The carbon-based chemistry is fairly well understood, but  the oxygen chemistry is not.
Two molecules vital  to the understanding of the oxygen chemistry are water and molecular oxygen, since some
chemical models predict that  a large fraction of oxygen could be in the form of these molecules.  

Observations of the ground-state rotational water transition with  NASA's Submillimeter Wave Astronomy Satellite \citep[SWAS;][]{2000ApJ...539L..77MelnickSWAS}
and the Odin satellite  \citep{2003A&A.402.21.Nordh.etal}   have both shown  very low abundances of molecular oxygen, 5\x10$^{-8}$ \citep{2007A&A...466..999Larsson}, and water, \mbox{10$^{-8}$\,--\,10$^{-9}$} \citep{2003A&A...402L..47Henrik}, in the \emph{extended} cold clouds in contrast to the predictions from chemical models.  However,  at smaller scales
in  warm or shocked regions and  outflows,  where the density is higher than 10$^{5}$\,--\,10$^{6}$\,\cmcub,
SWAS \citep[e.g.][]{2000ApJ...539L..87M} and Odin \citep[e.g.][]{2007A&A...476..807PaperII}
have also shown   \emph{locally}  increased water abundance
by several  orders of magnitude, 10$^{-6}$\,--\,10$^{-4}$ with respect to molecular hydrogen.
The  analysis of the data is, however, somewhat difficult and uncertain due to observations of a single line and  large beam-widths, but is in agreement with
observations of a wealth of highly excited water vapor transitions with ISO 
\citep[the Infrared Space Observatory,][]{2005SSRv..119...29Cernicharo}. 
These locally high abundances support the predictions that water should be an
important cooling agent in the star-forming process, especially  at high temperatures and densities   \citep[e.g.][]{1995ApJS..100..132Neufeld, 2003ApJ...582..830B}.
The strong dependence  of the  water chemistry     on  the temperature changes seen in star-forming regions also make
water  an excellent tracer of shocks and outflows from  protostars   still   hidden within their dusty envelopes \citep[e.g.][]{1998ApJ...499..777B}.

The intention of this paper is to continue the previous work with  ISO, SWAS, and Odin, and also to be a valuable input to future observations with the
Heterodyne Instrument for the Far Infrared (HIFI) on board the Herschel\footnote{http://herschel.esac.esa.int/} satellite to be launched in  2009. Our aim is to investigate how 
the water abundance depends on the environment in a star-forming region. 
The high spectral resolution (0.5\,\kms) and 
the smaller beam-size of Odin compared to SWAS (2$\farcm$1 vs.   3$\farcm$3\x4$\farcm$5), which enable higher spatial resolution maps,  allows a more detailed analysis of the line profile observed at each map position. 
We can also obtain an 
estimate of the beam-filling   by comparing  SWAS and Odin line profiles 
if the same object has been observed.
If not properly taken into account
the beam-filling may otherwise
influence the results by orders of magnitude.

In addition to the water observations
a second goal of our work is to perform and analyse observations of the important   ground-state rotational transition of ammonia.
While ammonia has been extensively observed in the inversion transitions at cm-wavelengths  \citep[e.g.][]{1986A&A.157.207.Ungerechts.etal, 1993ApJ.417.613.Zhou.etal}, the 1$_0$\,--\,0$_0$ transition, however, cannot be observed from ground due to the opaque terrestrial atmosphere.
The first  observation of this transition was performed by
the Kuiper Airborne Observatory  \citep{1983ApJ...271L..27Keene}, and was later followed up with Odin  \citep{2003A&A...402L..69Larsson, 2003A&A...402L..73Liseau,  2007A&A...476..791O}. 

Our chosen target is the
well known molecular cloud and bright nebula Sh 2-140, more
commonly known as S140  \citep{1959ApJS....4..257Sharpless, 1978ApJ...219..896Blair.etal}. It is  located at a distance of 910 pc \citep{1974PDAO...14..283Crampton}  in the large \mbox{L1202/L1204} dark cloud. This region is illuminated from the south-western side by the nearby  \mbox{B0.5V} star  HD\,211880, creating a bright HII region and an almost edge-on PDR adjacent to the molecular cloud and extending deeply into the  cloud.   
The core in S140 also contains a small cluster of deeply embedded young stellar objects which illuminate the cloud from within. One or more of these early B-stars are undergoing some mass-loss process and     molecular outflows are observed from this   dense and warm region \citep[e.g.][]{1977ApJ...213L..35Rouan.etal, 1987ApJ...312..327Hayashi, 1993A&A...277..595M.Minchin.etal, 2002A&A...383..540Preibisch}. 
The brightest infrared source in S140, called IRS1,  is on small scales composed of several sources,  two of which might be powering   bipolar molecular outflows in almost perpendicular directions \citep{2007AJ....134.1870T.Trinidad.etal}. 
The increased temperature and density in both the PDR and the outflows are likely expected to produce a higher than average water abundance. For these reasons, S140 is a promising target for our purposes. 

\begin{figure}[h] 
  \resizebox{\hsize}{!}{\includegraphics{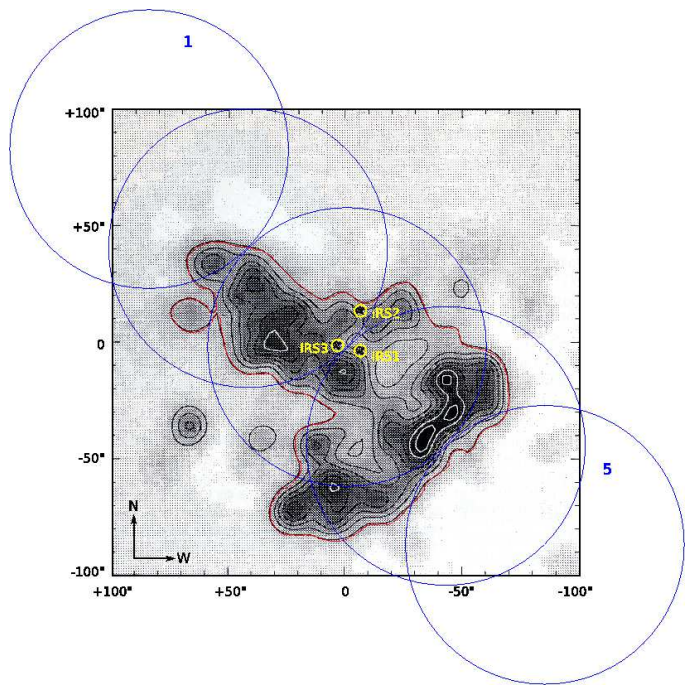}}
 \caption{A CI map (492\,GHz) of S140 observed with JCMT \citep{1993A&A...277..595M.Minchin.etal}.
The Odin 2$\farcm$1 beam at the five strip positions is also shown. Position 1 in NE to position 5 in SW.}
 \label{h2o_no2 strip on CI map}
\end{figure}

Both  ISO  \citep{2003A&A...403.1003Boonmanb}  and SWAS \citep{2000ApJ.539.101.Snell.etal,  2000ApJ.539.119.Ashby, 2000ApJ.539.115.Ashby, 2008ApJ...674.1015F} have observed water in S140 and several attempts have been made to model these   observations  \citep[e.g.][]{ 2000ApJ.539.119.Ashby, 2000ApJ.539.115.Ashby, 2003A&A...406..937Boonman, 2005A&A.440.559.Poelman.Spaans, 2006A&A.453.615.Poelman.Spaans}. However, they   all have had difficulties to explain  the unusual, narrow 1$_{1,0}$\,--\,1$_{0,1}$ emission line  with no or little evidence of self-absorption. 
In order to obtain more information about the   water emission
we have conducted
strip map observations of S140, starting   deep within the molecular cloud and reaching outside the bright rim (see Fig.~\ref{h2o_no2 strip on CI map}).
Strip maps are also simultaneously performed of the  previously unobserved \nh3(1$_0$\,--\,0$_0$) transition.

To aid the analysis of water and ammonia, we have  in addition performed  \water18 observations in the central position,
strip maps of \mbox{CO\,(5\,--\,4)} and \mbox{\13co\,(5\,--\,4)} with the Odin satellite, and a
\mbox{\13co\,(1\,--\,0)} map with the Onsala 20-m telescope.

\section {Observations} \label{Section Observations}

All transitions, except  \13co(1\,--\,0),
were acquired with the Odin satellite   from 2003 to 2006 with a total of 561 orbits (see Table~\ref{table overview frequencies}). The
calibration procedure (the chopper wheel method) is described in \citet{2003A&A.402.35O.Olberg.etal}.
The Odin \mbox{1.1 m} offset Gregorian telescope has a circular beam at 557\,GHz with a Full Width of Half Maximum (FWHM) of  2$\farcm$1 \citep{2003A&A.402.27.Frisk.etal}.
Being outside the atmosphere, and with an exceptionally high
main beam efficiency, $\eta_{\mathrm{mb}}$\,=\,0.9, our  intensity  calibration is very accurate.  
The intensity scale in the figures is expressed in terms of antenna temperature $T^*_\mathrm{A}$. In all calculations, however,   the main beam efficiency is properly taken into account.
The reconstructed pointing offset was $<$15$\arcsec$~during most of the time.
The simultaneous observations of \h2o and \nh3, and     of  \13co, CO and \h2o
 guarantee their pointing  to be identical.
 
Three different  tunable submm receivers were used, having average single-sideband (SSB) system temperatures of about 3\,300\,K.
They were used in combination with a hybrid autocorrelator spectrometer (300\,--\,400\,MHz working bandwidth) and an acousto-optical spectrometer (1\,GHz working bandwidth), both with a channel spacing of $\sim$0.3\,\kms.
All observations were performed in the sky-switching mode with reference sky beams at 42$^\circ$ distance and 4.4$^\circ$ FWHM. To correct for ripple, observations of a reference off position 30\arcmin~east of S140 were conducted as well.  


\begin{table} [!h]
\caption{Observed molecular transitions. All lines are observed with the Odin satellite except the  \13co(1\,--\,0) map  which is observed with the Onsala 20-m telescope.
}
\label{table overview frequencies} 
\centering
\begin{tabular} {ll r r l r}
 \noalign{\smallskip}
\hline
\hline
 \noalign{\smallskip}
Species &Transition&Freq. & $E_\mathrm{u}^a$&$n_\mathrm{crit}^b$&Int. time     \\
&&[MHz]&[K]& [\cmcub]& [h]\\
 \noalign{\smallskip}
\hline
 \noalign{\smallskip}
\h2o	 	 &  1$_{1,0}$\,--\,1$_{0,1}$     &  556\,936.0 &61&3\x10$^{8\,\,c}$  &     41   \\
\water18 &  1$_{1,0}$\,--\,1$_{0,1}$    & 547\,676.4  &60& 3\x10$^{8\,\,c}$       &      91\\
CO &   5\,--\,4    &  576\,267.9   &83& 2\x10$^{5\,\,c}$  &     9 \\
\13co &   5\,--\,4    &  550\,926.3 & 79   &2\x10$^5$   &   9    \\
\13co &   1\,--\,0    &   110\,210.4    &5& 7\x10$^2$  & 2 min/pos      \\
\nh3 &    1$_0$\,--\,0$_{\,0}$   &    572\,498.1   &29& 5\x10$^{7\,\,c}$  &  16 \\	
 \noalign{\smallskip}
\hline
\end{tabular}
\begin{list}{}{}
\item$^{{a}}$Upper state energy.
\item$^{{b}}$Critical density calculated for a temperature of 40\,K.
\item$^{{c}}$Can be considerably reduced because of radiative trapping.
\end{list}
\end{table}


The five point strip maps have 60$\arcsec$~step size with
strip centre coordinates  \mbox{R.A. 22$\fh$17$\fm$42\,$\fs$0}, \mbox{Dec. $+$63$\fdg$03$\farcm$45$\farcs$0} (B1950) at the cloud core (our position 3, Fig.~\ref{h2o_no2 strip on CI map}).
Position 1 is in NE and position 5 in SW.
The  frequency scale is set   relative to a source LSR velocity of $-$7.5\,\kms. 

In addition, the Onsala 20-m telescope was used in  2006  for a \mbox{\13co(1\,--\,0)}  10\x11 point map of the S140 molecular cloud with a spacing of 30\arcsec. 
The SSB system temperature was $\sim$500\,K, the main-beam efficiency\,=\,0.5, and the FWHM beam at 110.201\,GHz is 34\arcsec.

\section {Results} \label{Section results}

\begin{figure}[h] 
\includegraphics[scale=0.5]{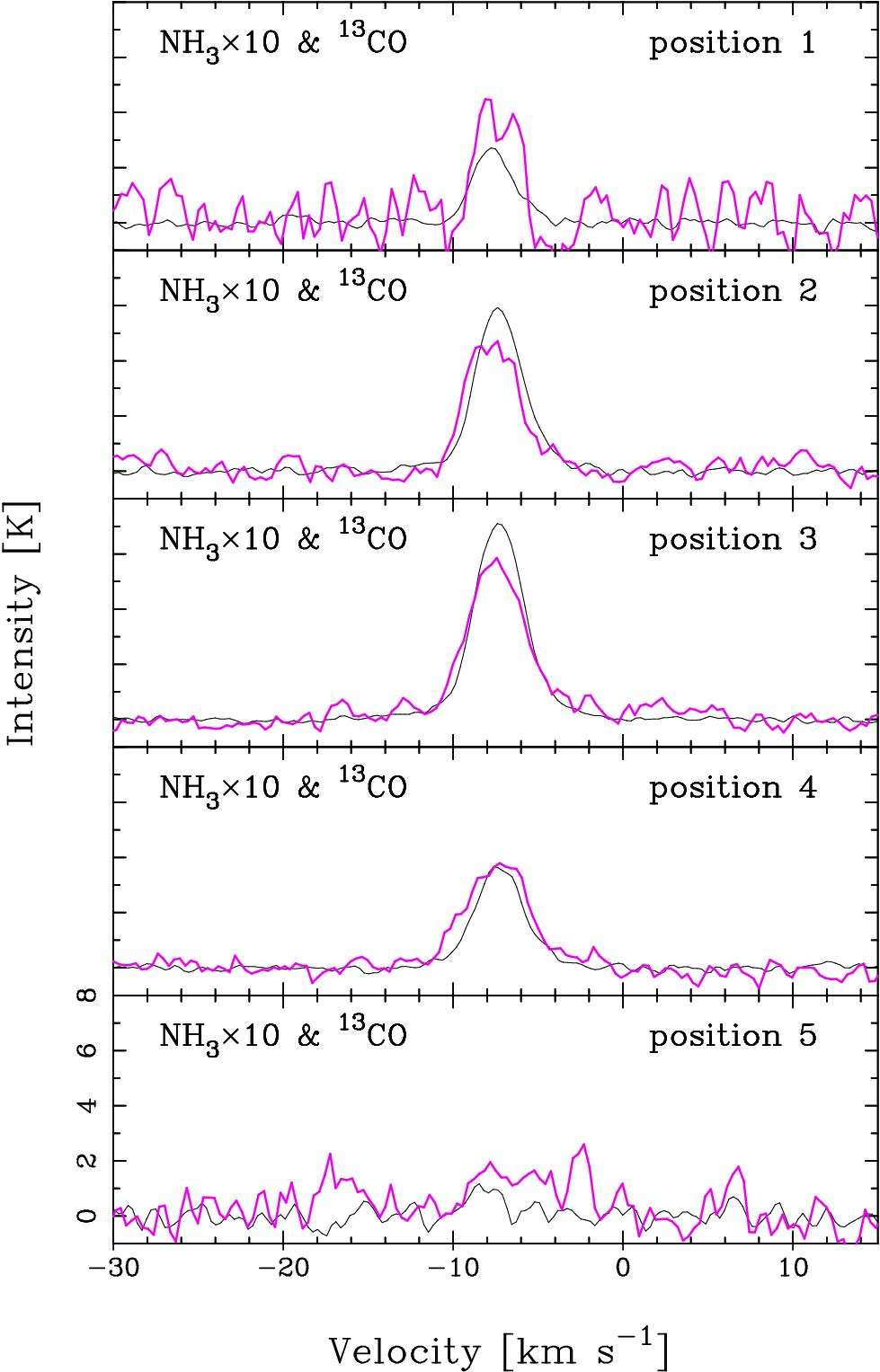}
 \caption{Comparison of the \nh3 1$_0$\,--\,0$_{\,0}$ (thick magenta line) and \13co(5\,--\,4) (thin black line) line profiles in a strip map   from NE to SW across the PDR.}
 \label{c-13o and nh3 no2 strip comparison}
\end{figure}

\begin{figure}[h] 
\includegraphics[scale=0.5]{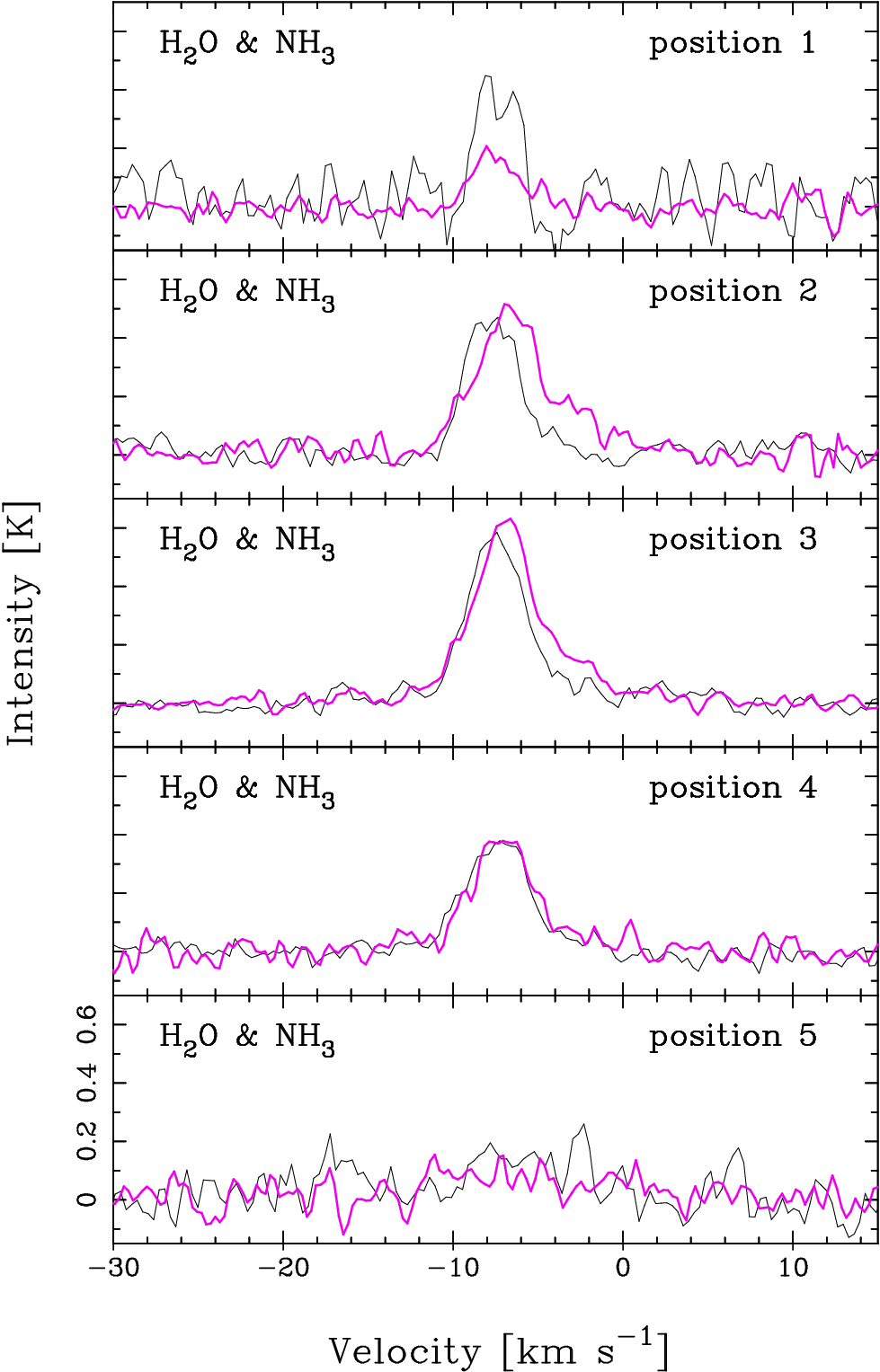}
 \caption{Comparison of the \h2o (thick magenta line) \trans~and the \nh3 (thin black line) 1$_0$\,--\,0$_{\,0}$  line profiles in a strip map from NE to SW across the PDR.}
 \label{nh3 and h2o no2 strip comparison}
\end{figure}

The observed Odin spectra  of CO,  \h2o, and \nh3 are shown together with 
\13co(5\,--\,4) in Figs.~\ref{co and c-13o comparison strip no2}\,--\,\ref{c-13o and nh3 no2 strip comparison} to allow a comparison of the line profiles of the different molecules at each position.
The water and ammonia spectra are also shown together in Fig.~\ref{nh3 and h2o no2 strip comparison}.
No \water18 emission was detected in the central position
(Fig.~\ref{spectrum h2o18}, on-line material)  at  an rms noise level of   8\,mK.
Tables with 1$\sigma$ rms,  integrated intensities, peak antenna temperatures,  line widths and amplitudes of Gaussian fits, and centre velocities 
for   species observed with Odin are found in the on-line Tables~\mbox{\ref{Odin table}\,--\,\ref{result_tablenh3}.}  
The on-line material also includes figures of 
Gaussian fits to \mbox{\13co(1\,--\,0)} and $J$\,=\,5\,--\,4,  \h2o and  \mbox{\nh3(1$_1$\,--\,0$_{\,0}$)}   \mbox{(Figs.~\ref{13co 5-4 center 2 Gauss}\,--\,\ref {nh3 center 2 Gauss})}.

Below we will discuss the observations and some straight forward results, while the  model results for the narrow PDR component are presented in
Sect.~\ref{subsection RADEX} and \ref{subsection Spaans model}.


\subsection{Carbon monoxide} \label{simple CO results}

\begin{figure}[h] 
\includegraphics[scale=0.5]{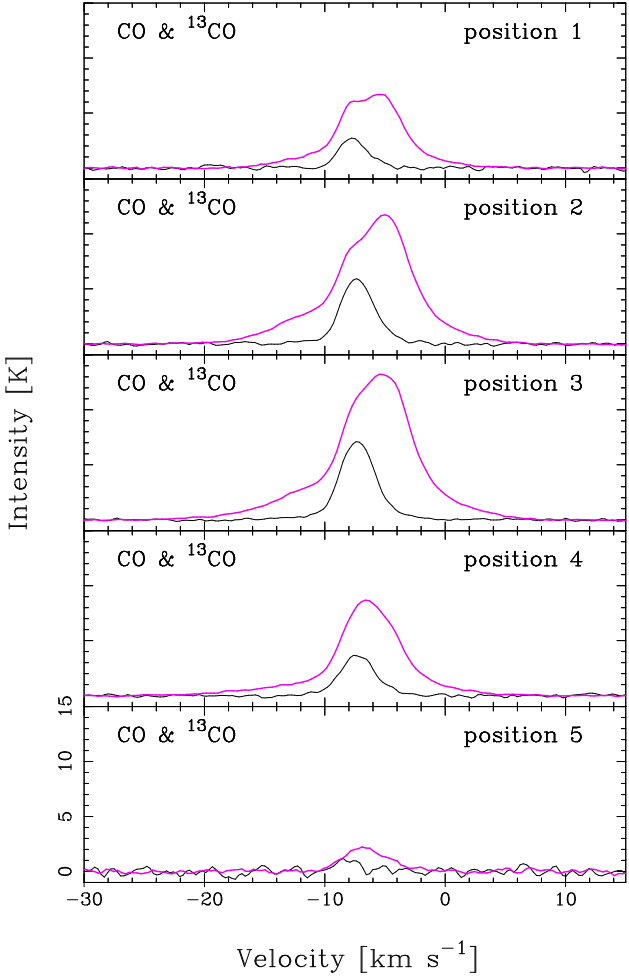} 
 \caption{Comparison of the CO(5\,--\,4) (thick magenta line) and \mbox{\13co(5\,--\,4)} (thin black line) line profiles in a strip map from NE to SW across the PDR.}
 \label{co and c-13o comparison strip no2}
\end{figure}

The CO(5\,--\,4) and \13co(5\,--\,4)  observations show a narrow   line 
from the PDR in all positions superimposed on a   broader outflow feature seen in CO at positions \mbox{1\,--\,4} 
and in \13co  
at the centre position
(Fig.~\ref{co and c-13o comparison strip no2}). 
The emission peaks about 20\arcsec~from the central position towards position 2 deeper in the cloud.
A comparison of the $^{12}$CO and \13co line shapes suggests strong CO(5\,--\,4) self-absorption by  lower excitation foreground gas.
This is also discussed by  e.g. \citet{1993A&A...277..595M.Minchin.etal} and \citet{2003A&A...406..937Boonman} where
the higher $J$-level CO lines are strongly self-absorbed. No strong self-absorption is visible in \13co.

The mean opacity of $^{12}$CO and \13co
can be determined by 
comparison of their antenna peak temperatures if the  isotopologue abundance ratio $R$ is known. Assuming that  
the opacity of $^{12}$CO is larger than the opacity of \13co  by a factor      $R$, we will have
\begin{equation}\label{tau_iso_comp}
\frac{T^*_\mathrm{A, 13}}{T^*_\mathrm{A, 12}}
 = \frac {J(T_{\mathrm{ex}}^{13})\,(1-e^{-\tau_{13}}) \, \eta_{\mathrm{\,bf, {13}}}}{J(T_{\mathrm{ex}}^{12}) \, (1-e^{-\tau_{12}}) \, \eta_{\mathrm{\,bf, 12}}}= \frac {J(T_{\mathrm{ex}}^{13})\, (1-e^{-\tau_{13}}) \,
 \eta_{\mathrm{\,bf, 13}} }{J(T_{\mathrm{ex}}^{12})\, (1-e^{-R\,\tau_{13}})  \, \eta_{\mathrm{\,bf, 12}}},
\end{equation}
where the
radiation temperature, $J(T_\mathrm{ex}$), is a function of the excitation temperature, $T_\mathrm{ex}$,
\begin{equation} 
J(T_\mathrm{ex})  = \frac{h \nu}{k} \frac{1}{e^{\,h\nu/kT_\mathrm{ex}}-1}.
\end{equation}

The beam-filling factor
$\eta_{\mathrm{bf}} =\theta_\mathrm{s}^2 / (\theta_\mathrm{s}^2+\theta_\mathrm{mb}^2)$, assuming that both the source brightness distribution and the antenna response are circularly symmetric and Gaussian. 

Assuming that the excitation temperatures are the same  for both species,  co-spatial   emissions  and using $R$\,=\,70,
the opacities of the narrow   component of $^{12}$CO  and \13co are estimated  to be $\lesssim$\,70 and $\lesssim$1, while the opacities in the broad outflow component are $\lesssim$\,7 and $\lesssim$\,0.1, respectively. Due to the self-absorption of $^{12}$CO the opacities are upper limits.

The size of the emission region can be estimated by comparing Odin and SWAS  antenna temperatures
\begin{equation} \label{source size SWAS/Odin}
\frac{T_\mathrm{A, Odin}^*}{T_\mathrm{A, SWAS}^*}=\frac{(3\farcm3 \times60\times4\farcm5 \times60)+\theta^2}{(2\farcm1 \times60)^2+\theta^2},
\end{equation}
where $\theta$ is the effective circular Gaussian source size in arcseconds.
For the central position the ratio of the Odin and SWAS \13co peak antenna temperatures is 
$\sim$2.5 which is equivalent to a mean source size of $\sim$100$\arcsec$. 

A second approach to estimate the source size is performed 
by plotting   the antenna temperatures in our Odin strip map vs. position, and fitting  a Gaussian. After deconvolution with the 126\arcsec~Odin beam, the resulting one dimensional source size is about 106\arcsec. Together with the mean source size of $\sim$100$\arcsec$, this implies a relatively circular source.

The excitation temperature can be estimated by plotting a
rotation diagram  of several transitions. 
The level populations are  given by the Boltzmann equation and
the excitation temperatures for all the energy levels are assumed to be the same.
The beam-filling corrected   integrated intensities of all lines  are plotted as a function of  the   upper state energy in a semi-log plot
based upon the relation \citep{2007A&A...476..807PaperII}
\begin{equation}\label{full solution for rotation diagram}
\ln \, \frac {8 \pi\,k \,\nu_{{ul}}^2}{h\,c^3 g_{u}\,A_{ {ul}}}  \int {T_{\mathrm{b}}}  \, \mathrm{d} \upsilon = \ln \, \frac{N_{\mathrm{ROT}}}{Q(T)} - \frac {E_{u}}{kT_\mathrm{ex}}.
\end{equation}
where
$k$ is the Boltzmann constant, $\nu_{{ul}}$ is the frequency of the transition, $h$ is the Planck constant, $c$ is the speed of light,  $A_{{ul}}$ is the Einstein $A$-coefficient for the transition,   $T_\mathrm{b}$ is the  brightness temperature (observed antenna temperature corrected for beam-efficiency and beam-filling),
$g_u$ and $E_u$ are the statistical weight and energy of the upper state, respectively, $Q(T)$ is the partition function,   $T_\mathrm{ex}$ is the excitation temperature for the transition, and $N_{\mathrm{ROT}}$ is the total column density of the species obtained from the rotational diagram.
As customary the frequency axis $\nu$ has been converted to a 
velocity axis $\upsilon$ using the speed of light.
A  least-squares fit to the data will  produce a straight line with slope $-1/T_{\mathrm{ROT}}$. 
If we extrapolate the line to $E_{u}$\,=\,0\,K,  the total column density is found from the intersection  of the y-axis.

\begin{figure}[h] 
\includegraphics[scale=0.5]{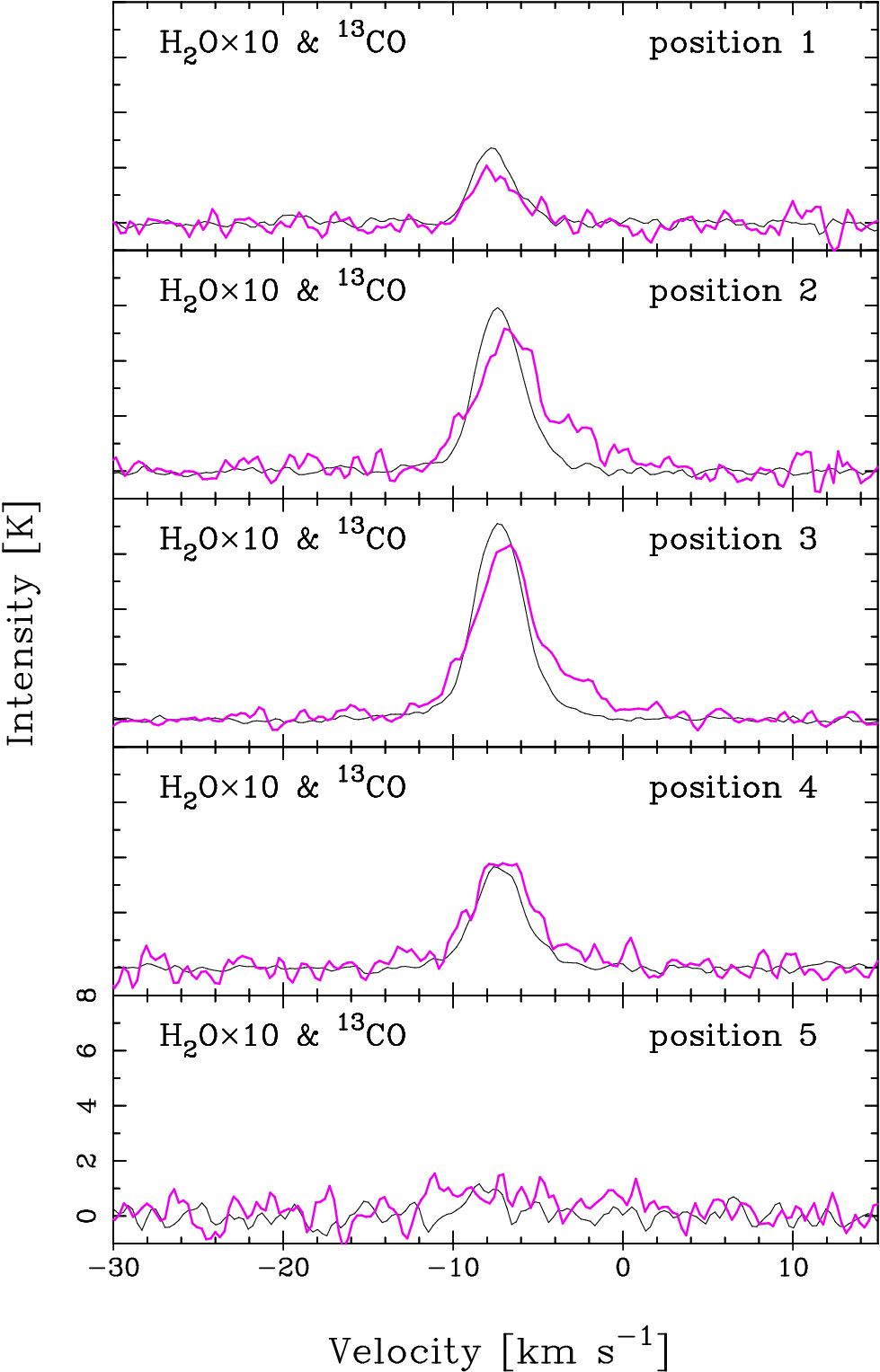} 
 \caption{Comparison of the H$_2$O \trans~(thick magenta line) and \13co(5\,--\,4) (thin black line)  line profiles in a strip map  from NE to SW across the PDR.}
 \label{h2o and c-13o strip comparison no2}
\end{figure}

Our rotation diagram of the    broad outflow \13co component  in the central position (on-line Fig.~\ref{Rotdiagram 13co outflow}), including the Onsala 20-m $J$\,=\,1\,--\,0 transition, the \mbox{$J$\,=\,2\,--\,1} and 3\,--\,2 transitions   from \citet{1993A&A...277..595M.Minchin.etal}, and the Odin $J$\,=\,5\,--\,4  transition, produces $T_\mathrm{ROT}$\,=\,24$\pm$2\,K and $N_\mathrm{ROT}$\,=\,(2.5$\pm$0.4)\x10$^{16}$\,\cmsq. All antenna temperatures are corrected with respective beam-filling and beam-efficiencies.
This temperature is in agreement with the previously estimated lower limit of the outflow excitation temperature of  about 15\,--\,20\,K \citep{1993A&A...277..595M.Minchin.etal}. The low temperature of the outflow  indicates that the emission arises from gas far behind the shock front where the gas has already cooled \citep{1984ApJ...284..176S}. 

The excitation temperature of 
the  narrow  \13co PDR emission in the central position is also estimated by means of a rotation diagram
(on-line Fig.~\ref{Rotdiagram 13co 4 lines PDR}), including the above mentioned transitions with addition of the
$J$\,=\,6\,--\,5 line  from \citet{1993ApJ...405..249Graf},   but excluding  $J$\,=\,1\,--\,0  which seems to   originate only in the colder gas component.
This produces
$T_\mathrm{ROT}$\,=\,69$\pm$27\,K,  
in agreement with \citet{1993A&A...277..595M.Minchin.etal} who find a
PDR excitation temperature of about 70\,K at IRS  and a steep rise in temperature to about 250\,K at
the dissociation front \mbox{60\arcsec\,--\,80\arcsec} SW of IRS1.
Our PDR  column density, $N_\mathrm{ROT}$(\13co)\,=\,(3.2$\pm$1.8)\x10$^{16}$\,\cmsq, is relatively uncertain but 
in agreement with the {\tt RADEX} result in Sect.~\ref{subsection RADEX}, $N$(\13co)\,=\,2.8\x10$^{16}$\,\cmsq,
and with \citet{1994ApJ...425L..49P}, 3.4\x10$^{16}$\,\cmsq.



\subsection{Water} \label{subsection water}

The \emph{ortho}-water 1$_{1,0}$\,--\,1$_{0,1}$ ground-state rotational transition  is detected in all positions except  outside the bright rim (position~5). 
Similar to CO and \13co, a
narrow  component (\deltav~$\sim$3\,\kms) superimposed on a broader outflow component (\deltav~$\sim$8.5\,\kms) is observed with emission originating around the central position. Like CO,
the water line profile is   dominated by outflow emission,
clearly seen in a fit of two Gaussian components in the on-line Fig.~\ref{h2o center 2 Gauss}.
The  integrated intensity of outflow emission is only about 20\% for \13co compared to 60\% for water which in addition
mainly shows a red wing  
(Fig.~\ref{h2o and c-13o strip comparison no2}). This apparent strong outflow emission may not, however, reflect the true relative abundances of the PDR and outflow components if the line is optically thick. 
The opacity is several times higher   in the PDR  than in the outflow and,  depending on the opacity, this may lower the fraction of outflow emission.
  The \13co emission is on the other hand almost optically thin which gives   more accurate relative amounts of the respective emissions.

The 
non-detection of \water18 can be used to set upper limits on the opacity of the main water line  using Eq.~\ref{tau_iso_comp}. 
However, the calculation is not as  straightforward as for CO.
Because of the high critical density   (the ratio of the spontaneous de-excitation and   collision coefficients) of the \h2o 557\,GHz line, the emission of this transition   is highly sub-thermal and has a  high opacity even for rather low abundances. 
This high optical depth  allows the excitation
temperature to be enhanced by radiative trapping. As a result, the 
corresponding lines of H$_2$O and H$_2^{18}$O  have quite different 
values of excitation temperature. Their excitation temperatures will also be much lower than the kinetic temperature.
Using the {\tt RADEX} code in Sect.~\ref{subsection RADEX} for a kinetic temperature of 55 K and a molecular
hydrogen density $n({\rm H}_2)$\,=\,4\x10$^5$\,cm$^{-3}$, we find excitation
temperatures of 8.2 and 6.3\,K for the $1_{1,0}-1_{0,1}$ transitions of
\h2o and H$_2^{18}$O, respectively. If the abundance ratio
[$^{16}$O/$^{18}$O]\,=\,330 in water \citep{2007A&A...476..807PaperII} and if the two
isotopologues are distributed similarly, then the 5$\sigma$ upper limit on
the intensity of the H$_2^{18}$O line implies that the optical depth of
the H$_2$O line is $\lesssim 60$.
If the assumed density is lower by a factor of 10, the 
difference in excitation temperatures will increase  and change the 
upper limit  to  $\lesssim$120.

In addition, the
high \h2o opacity often  produces   self-absorbed spectra. 
Our observations, however, do not
show any obvious signs of self-absorption,  in contrast to our  CO  lines. 
When comparing the centre velocities of the narrow components of water and CO   to   \13co,  we find that the velocities of     water and \13co are well matched (Fig.~\ref{h2o and c-13o strip comparison no2} and Table~\ref{result_tableCO}\,--\,\ref{result_tableh2o}), while  CO and \13co in positions 1\,--\,4 are not (Fig.~\ref{co and c-13o comparison strip no2}). This supports 
our conclusion
that there is no severe self-absorption in \h2o in contrast to 
observations of other sources in which corresponding
self-absorptions were seen in CO(7-6) and in ISO measurements of H$_2$O  \citep{2003A&A...406..937Boonman}.

The lack of self-absorption  points to a  rather constant excitation across the region \citep[cf.][]{1984ApJ...276..625S}.
It  also  implies a lower water abundance in the blue than in  the red outflow.
A higher water  abundance in the red wing  may be caused by outflowing gas pushing into the ambient high density cloud   thereby producing water. The blue outflow,  on the other hand,  is leaving the molecular cloud, and  expands into an ionized region with low density gas with no or very low water production as a result. 
This is in agreement with the results in \citet{2008ApJ...674.1015F} who investigated water abundance in molecular outflows with SWAS
and find about five  times higher abundance in the red   outflow  compared to the blue.

The sizes of the emitting regions are estimated in the same way as for \13co (Eq. \ref{source size SWAS/Odin}).
For the central position the ratio of the Odin and SWAS peak antenna temperatures  is
$\sim$2.4, equivalent to a mean source size of $\sim$100$\arcsec$. 
This   confirms previous suggestions that the  water emission   does not fill the SWAS beam
\citep{2003A&A...406..937Boonman}.
To obtain estimates of the size of each component,
we use 
the amplitudes of the fitted Gaussians from the Odin and SWAS spectra. The ratios of the PDR and outflow components are
$\sim$2.2 and 
$\sim$2.6, which corresponds to mean source sizes of 120\arcsec~and 85\arcsec, respectively.

Gaussian fits to the amplitudes in each position in the Odin strip map   results in one-dimensional source sizes of  $\gtrsim$150\arcsec~and
$\lesssim$50\arcsec~for the PDR and outflow, respectively.
Taken together, 
these two calculations of source sizes
suggest that the water emission from the PDR is
more extended in the NE\,--\,SW direction (same direction as our strip map), while the contrary applies to the outflow, indicating a NW\,--\,SE elongation. 

A third approach to estimate the beam-filling can be achieved by using the    radiative transfer equation. 
With a constant source function, the solution of the radiative transport equation is  \mbox{\citep[cf.][]{2007A&A...476..807PaperII}}

\begin{equation}\label{solution}
T^*_\mathrm{A} =T_\mathrm{b}  \, \eta_{\mathrm{mb}}\, \eta_{\mathrm{bf}}=(J(T_\mathrm{ex})-J(T_\mathrm{cont}))  \, (1-e^{-\tau})  \, \eta_{\mathrm{mb}}\, \eta_{\mathrm{bf}}, 
\end{equation}
where $J(T_\mathrm{cont})$ is the background continuum radiation.

For an optically thick line and neglecting the background radiation
\begin{equation}\label{optically thick line solution}
T^*_\mathrm{A}  \approx   J(T_\mathrm{ex}) \eta_{\mathrm{mb}}\, \eta_{\mathrm{bf}}.
\end{equation}
If $T_\mathrm{ex}$  is known, then the beam-filling can be estimated.
Using   $T_\mathrm{ex}$\,=\,8.2\,K obtained from {\tt RADEX} and Equation~\ref{optically thick line solution} give
a beam-filling factor that corresponds to a 
PDR source size of about 110\arcsec, which is close to the source size of 120\arcsec~obtained from the relative temperatures from Odin and SWAS.
This implies that the gas is  somewhat clumped, and that the surface filling fraction of the high density PDR gas is $\sim$84\%. 
The clumpy structure of the cloud is   confirmed by previous observations \citep{1984ApJ...276..625S,1984ApJ...284..176S, 1993ApJ.417.613.Zhou.etal,1993A&A...277..595M.Minchin.etal,1992PASJ...44..391Hayashi, 2000ApJ.539.119.Ashby} and 
our model results in Sect.~\ref{subsection Spaans model}.
The amount of surface clumping is, however, not possible to determine very accurately  due to the 
uncertainty of the excitation temperature  which is affected by the assumed density. 
With $T_\mathrm{ex}$\,=\,9\,K and 10\,K the same calculations give source sizes of~86\arcsec~and 69\arcsec, with corresponding surface filling fractions of 50\% and 33\%, respectively. Including the background radiation in Eq.~\ref{solution} also introduces additional uncertainties.

\subsection{Ammonia} \label{subsection ammonia}

A complex energy level diagram with a wealth of transitions covering different temperatures and densities has made ammonia a very valuable diagnostic of physical conditions in the interstellar medium.
The 1$_0$\,--\,0$_{\,0}$ ground-state rotational transition, observed by Odin, has an upper state energy comparable to the extensively observed  inversion transitions \citep[e.g.][]{1986A&A.157.207.Ungerechts.etal, 1993ApJ.417.613.Zhou.etal}, but the critical density is about four orders of magnitude larger. It is therefore likely that these transitions partly probe different gas components. As for water,
the 1$_0$\,--\,0$_{\,0}$ transition has a high opacity and is highly sub-thermally excited in most cases. 

The 1$_0$\,--\,0$_{\,0}$  transition of \emph{ortho}-\nh3  is detected in all positions except outside the bright rim (position~5). The line profiles and centre velocities are  very similar to those of  \13co(5\,--\,4) with an exception in position~1 (Fig.~\ref{c-13o and nh3 no2 strip comparison}).
Our comparison of  ammonia and water  in Fig.~\ref{nh3 and h2o no2 strip comparison} shows an almost identical line profile at the bright rim (position~4), which suggests emission from the same gas and velocity fields at this position originating in the PDR.
In the central position  the water emission shows  a more pronounced outflow in the red wing than ammonia, although very similar in the blue outflow. 
The emission fraction of the outflow is 35\% for ammonia, while it is almost twice as high for water. 
This
suggests that
the outflow production mechanism is more efficient for water than ammonia.
The water emission from the outflow seems to be almost the same in position~2 and 3, while the ammonia emission from the outflow decreases in position~2. Deepest in the molecular cloud   a very weak water line is seen.  The ammonia emission is, however, considerably stronger than the water emission at this position.
This confirms previous observations of \nh3 inversion transitions \citep{1993ApJ.417.613.Zhou.etal, 1986A&A.157.207.Ungerechts.etal} who find a \nh3 peak further into the cloud at about 1\arcmin~northeast of IRS1.
When we plot the peak antenna temperatures vs. position  it is   obvious that we see additional emission in position~1. A Gaussian fit to positions \mbox{2\,--\,4} gives a source size of $\sim$130\arcsec, with a peak about 10\arcsec~into the cloud from the centre. This size is relatively uncertain due to a fit to only three positions, but agrees well with the source sizes estimated from \13co and \h2o.

The Odin observations indicate that in addition to the warm, dense PDR and a weak outflow around IRS, 
we also observe \nh3 from the same cold  and extended   gas deep in the cloud (position 1) as \13co(1\,--\,0) at a temperature about \mbox{20\,--\,30\,K}.
This is also in agreement with 
\citet{1986A&A.157.207.Ungerechts.etal} and \citet{1993ApJ.417.613.Zhou.etal} who  find a \nh3 rotation temperature of 40\,K in between our first and second position.  
Chemical models predict that \nh3 is not depleted in high density regions ($\lesssim$1\x10$^6$\,\cmcub) and is therefore relatively more abundant than other molecules such as water and CO \citep{1997ApJ...486..316BerginAndLanger, 2003cdsf.conf...63Bergin}.

The surface filling fraction of the narrow PDR component of ammonia is also estimated using the radiative transfer equation (Eq. \ref{optically thick line solution}). The excitation temperature is approximately the same as for water thereby producing similar surface filling factors. 
Together with
the similarity of the    line profiles, source sizes and  centre velocities of the \h2o, \nh3 and \13co(5\,--\,4) narrow components, this  points to emission from the same  high density PDR clumps with the same temperature of about 70\,K for all species \citep[cf.][]{2003A&A...406..937Boonman}.

 \begin{table*}
\caption{Observations and model calculations for the narrow PDR component in the {\tt RADEX} homogeneous model.$^a$}
   \label{Line intensities Radex}
    \centering
     \begin{tabular}{l l r r c c r c c r c}

     \hline\hline
     \noalign{\smallskip}
Species   &  Transition & Frequency  & \multicolumn{3}{c}{Observations}
 & Ref  &\multicolumn{4}{c}{Model calculations} \\
 & & $\nu$ & $T_{\rm b}$ & $4\pi I$ & $\Delta \upsilon$ & & $T_{\rm ex}$ &
 $\tau$ & $T_{\rm b}$ & $4\pi I$ \\
 & & [GHz] & [K] & [ergs s$^{-1}$ cm$^{-2}$] & [km s$^{-1}$] & & [K] &
 & [K] & [ergs s$^{-1}$ cm$^{-2}$] \\
\hline
 H$_2$O & \trans & 556.936 & 0.92 & \dots & 3.1 & $c$ &
  8.2 & 6.6 & 0.92 & $6.8\times 10^{-6}$ \\  
     \noalign{\smallskip}
 H$_2^{18}$O & \trans & 547.676 & $<0.051$ & \dots & (3.1)$^b$ & $c$&
  6.3 & 0.2 & $<0.05$ & $<3.5\times 10^{-7}$ \\ 
     \noalign{\smallskip}

 NH$_3$ &  1$_0$\,--\,0$_{\,0}$ & 572.498 & 1.1 & \dots & 3.3 &$c$ & 
  9.0 & 6.1 & 1.1 & $9.2\times 10^{-6}$ \\
  & (1,\,1) & 23.695 & 2.9 & \dots & \dots &$d$ &
  54 & 0.02 & 0.87 & $2.5\times 10^{-10}$ \\
  & (2,\,2) & 23.723 & 2.0 & \dots & \dots & $d$ &
  47 & 0.01 & 0.50 & $1.7\times 10^{-10}$ \\
   & (3,\,3) & 23.870 & 1.4 & \dots & \dots &$d$ &
  $-5.1$ & $-0.25$ & 2.2 & $9.0\times 10^{-10}$ \\
 & (4,4) & 24.139 & 0.15 & \dots & \dots & $d$ &  46 & 0.001 & 0.06 & $3.8\times 10^{-11}$ \\ 
     \noalign{\smallskip}

 $^{13}$CO & 2\,--\,1 & 220.399 & 12 & \dots & 3.3 & $e$&
  55 & 0.2 & 9.1 & $4.3\times 10^{-6}$ \\
  &  5\,--\,4 & 550.926 & 14.7 & \dots & 3.2 &$c$ &
  51 & 0.5 & 14.7 & $1.1\times 10^{-4}$ \\
     \noalign{\smallskip}

 CS & 2\,--\,1 & 97.981 & 8.6 & \dots & 3.0 & $f$ &
  47 & 0.2  & 9.1 & $3.5\times 10^{-7}$ \\
 & 3\,--\,2 & 146.969 & 10.2 & \dots & 3.0 & $f$ &
  20 & 1.0 & 10.0 & $1.3\times 10^{-6}$ \\
 &  5\,--\,4 & 244.936 & 5.4 & \dots & 3.0 & $f$ &
  13 & 1.3 & 5.5 & $3.3\times 10^{-6}$ \\
  &  6\,--\,5 & 293.912 & 3.3 & \dots & 2.2 & $f$ &
  12  & 0.7 & 3.1 & $3.3\times 10^{-6}$ \\
     \noalign{\smallskip}

 [C I] &  1\,--\,0 & 492.161 & 3.7 & \dots & 3.5 & $g$ & 
  54 & 0.1 & 3.7 & $2.1\times 10^{-5}$ \\
     \noalign{\smallskip}

 [C II] & 3/2\,--\,1/2 & 1900.537 & \dots & $4.4\times 10^{-3}$ & \dots
   & $h$ & 55 & 1.0 & 13.4 & $4.4\times 10^{-3}$ \\
     \noalign{\smallskip}

 [O I] & 1\,--\,2 & 4744.778 & \dots & $4.1\times 10^{-3}$ & \dots &
   $h$ & 46 & 0.9 & 0.88 & $4.5\times 10^{-3}$ \\

     \noalign{\smallskip}
     \hline
     \end{tabular}
\begin{list}{}{}
\item[$^a$]All results for $T_{\rm K}$\,=\,55\,K, $n({\rm H}_2)$\,=\,4\x10$^5$\,cm$^{-3}$. $T_{\rm b}$\,=\,Rayleigh-Jeans brightness temperature,  $4\pi I$\,=\,flux, $\Delta \upsilon$\,=\,line width, $T_{\rm ex}$\,=\,excitation temperature,
 $\tau$\,=\,line-center peak opacity. $^b$ From \h2o.
$^c$ This work. $^d$ \citet{1986A&A.157.207.Ungerechts.etal}. $^e$  \citet{1993A&A...277..595M.Minchin.etal}.   $^f$ \citet{1984ApJ...276..625S}. $^g$\citet{1994ApJ...425L..49P}. $^h$ \citet{1996A&A...315L.285E}.
\end{list}
   \end{table*}

 \subsection{Outflow abundances} \label{subsection Snell formula}

 The collisional de-excitation of  the ground-state rotational transition of \emph{ortho}-water is very low in the density ranges considered here due
to the
high critical density of $\sim$3\x10$^8$~\cmcub~at a temperature of 40\,K.
A simple analytic expression for the antenna temperature of a two-level system in the low collision rate limit  can therefore be used 
since collisional excitation always results in a photon that escapes the cloud.
Thus, even though our transition may be optically thick, it is effectively  thin \citep{1977ApJ...214...50L, 2000ApJ...539L..93S}. 
We use this approach to 
estimate  the water
abundance  in the outflow \citep{2000ApJ...539L..93S}
\begin{equation} \label{Snell}
\int{T_{\rm b}\,d\upsilon} = C\,n_{\mathrm{H}_2} \frac{c^3}{2 \nu^3 k}\, N(\mathrm{o-H_2O}) \frac{h \nu}{4\pi}\,\exp(-h\nu/kT_K),
\end{equation}
where $C$ is the collision rate which equals the product of the cross-section and  velocity.
Dividing Eq.~(\ref{Snell}) with $N$(H$_2$)  and scaling the calculated factor in  \citet{2000ApJ...539L..93S} to our derived rotational outflow   temperature of  24\,K (with $\sqrt T$) produces the fractional abundance of \emph{ortho}-water with respect to H$_2$ in the outflow
\begin{equation}\label{Snell formula}
X(o-\mathrm{H_2O})   = 3.2\times10^{19} \frac{\int{T_{\rm b}\,d\upsilon}}{N(\mathrm{H_2})\,n_{\mathrm{H}_2}}.
\end{equation}

The H$_2$ column density in the outflow is calculated from $N_\mathrm{ROT}$(\13co) of the broad component. Using  typical values of 
[CO/H$_2$]\,=\,1\x10$^{-4}$ and   [$^{12}$C/$^{13}$C]\,=\,70 we obtain $N(\mathrm{H_2})$\,=\,1.8\x10$^{22}$\,\cmsq. Together with    a range of mean molecular hydrogen densities of 5\x10$^{4}$\,--\,5\x10$^{5}$\,\cmcub~we obtain
a total water     outflow abundance of $\sim$\,2$\times10^{-8}$\,--\,2$\times10^{-7}$. This is in agreement with
\citet{2008ApJ...674.1015F} who found an outflow abundance in the blue wing of 2.4$\times10^{-8}$ and 1.0$\times10^{-7}$ in the red wing
using a kinetic temperature of 30\,K and a molecular density of 10$^5$\,\cmcub.

The ammonia abundance in the outflow  is estimated using 
the Local Thermal Equilibrium (LTE) approximation. 
Assuming LTE and adding corrections for opacity and beam-filling, the  total source averaged column density can be calculated as
\begin{equation} \label{NtotCorrected}
N_{\mathrm{LTE}}  =\frac{C_\tau}{\eta_{\mathrm{bf}}}\, \frac {8  \pi k \nu_{{ul}}^2}{h c^3} \frac{1}{ A_{{ul}}} \frac{Q(T)}{g_{{u}}}  \exp\,(E_{{u}}/kT_{\mathrm{ex}}) \int {T_{\mathrm{b}}}\,d \upsilon, 
\end{equation}
where 
$C_\tau$\,=\,$\tau/(1-\exp\,(-\tau))$  is  the opacity  correction factor.  
Using $C_\tau$\,=\,1, a
source size of  85\arcsec~to calculate  $\eta_{\mathrm{bf}}$  and an excitation temperature of 10\,K,
the  ammonia outflow column density is  estimated to be  
$\sim$1.2\x10$^{13}$\,\cmsq. This column density is thus not opacity-corrected and may therefore be underestimated. 
The ammonia outflow abundance  is calculated as
 $X$(\nh3)\,=\,$N_\mathrm{NH_3}$/$N_\mathrm{H_2}$ and is found to be $\sim$7\x10$^{-10}$.

\section{Model results:  RADEX -- a homogeneous model} \label{subsection RADEX}

In this section  
we use 
an enhanced version of the {{\tt RADEX}}\footnote{The published 
version will soon include these enhancements, see
{\tt http://www.sron.rug.nl/{$\sim$}vdtak/radex/index.shtml}} \citep{2007A&A...468..627Radex} code
to compute the intensities of the narrow components of
 the transitions observed by Odin, and  the intensities of
 several 
atomic and molecular transitions that previously have been observed in  S140. 
The {\tt RADEX} codex 
applies a very simple method of mean escape probabilities to the
radiative transfer and yields results that are similar to those in
the large-velocity-gradient approximation, assuming an isothermal and homogeneous medium, but without an implicit
assumption of a gradient in velocity. 
Although much more sophisticated
methods of radiative transfer have been used to construct models of the
atomic and molecular line emission from this region, it is useful to 
examine a simple, internally consistent model of a homogeneous cloud 
that reproduces the principal observed facts. Such a model can be used
in particular to assess the role of radiative coupling of molecular
excitation to the intense continuous radiation within the PDR.
In Sect.~\ref{subsection Spaans model}, these results are 
compared to the results from
a  3D inhomogeneous PDR model  that calculates the temperature  and abundance distributions, and a multi-zone escape probability method that calculates the level populations and  intensity distributions of water and ammonia  \citep{2005A&A.440.559.Poelman.Spaans, 2006A&A.453.615.Poelman.Spaans}. 

The molecular data files for H$_2$O
and NH$_3$ have been enlarged, compared with those previously available
through the Leiden Atomic and Molecular Database (LAMDA)\footnote{\tt 
http://www.strw.leidenuniv.nl/{$\sim$}moldata}. 
Data for the \emph{ortho} and \emph{para} forms of these molecules have been combined
into a single file and infrared transitions have been added. Collision
rates published by \citet{2007A&A...472.1029F} for H$_2$~+~H$_2$O and by
\citet{2004MNRAS.347..323F} for e$^-$~+~H$_2$O are now used.
Two different data
files have been used to analyse the excitation of NH$_3$. The smaller
includes only the lowest 112 levels of the vibrational ground state and
140 radiative transitions between them. A larger file, derived mainly
from the HITRAN\footnote{\tt http://cfa-www.harvard.edu/hitran} database, has also been tested: it contains 2\,392 levels
and 15\,067 radiative transitions, including lines of several $\nu_2$
and $\nu_4$ vibrational bands in the mid-infrared.

We   include 
a  simple dust model of the broad-band 
spectrum \citep{1983ApJ...271..625S,1995A&A...298..894M.Minchin.etal, Ney}  at submm and far-infrared wavelengths in order  to 
characterize the internal radiation sensed by
the molecules (a detailed description is found in Appendix~\ref{on-line dust model construction} of the on-line material).

The principal parameters needed to specify a model are the kinetic 
temperature $T_{\rm K}$ and the average number density of molecular hydrogen
$n({\rm H}_2)$.  
Our adopted dust model implies $N({\rm H}_2)$\,=\,4.7$\times 10^{22}$\,cm$^{-2}$
and an average density $n({\rm H}_2)$\,=\,3\x10$^4$\,cm$^{-3}$ 
in a homogeneous, spherical cloud
with a source
size $L$\,=\,0.5\,pc (derivation in Appendix~\ref{on-line dust model construction}, on-line material).
Unfortunately, this average density is inconsistent
with our observations.
Although a uniform {\tt RADEX} model can be constructed based
upon this density,
this low-density model  
cannot explain the observed intensities of the  pure rotational lines of H$_2$O
and NH$_3$ 
unless the line-center optical depths are of the order of 200 and
100, respectively.
Such large opacities would imply significant 
line broadening through saturation of the emission, which conflicts with the
observed narrow profiles.

High densities  of 7\x10$^5$\,\cmcub~are found by multitransition CS observations
\citep{1984ApJ...276..625S},
and 5\x10$^5$\,\cmcub~from \nh3 inversion transitions \citep{1986A&A.157.207.Ungerechts.etal}.
We include the CS transitions in our {\tt RADEX} model  in addition to our own observations, and our best fit suggests a mean molecular density of 4\x10$^5$\,\cmcub~and a temperature of 55\,K.
This suggests that
the strongly emitting molecules occupy only a  7\%  fraction of the volume. 
The H$_2$ column density, which is needed to obtain the  abundances, $X$(x)\,=\,$N$(x)/$N$(H$_2$), is calculated from the $N$(\13co) obtained with  {\tt RADEX}. This column density is also in agreement with $N_\mathrm{ROT}$(\13co) found in Sect.~\ref{simple CO results}.
Using [$^{12}$C/$^{13}$C]\,=\,70 and [CO/H$_2$]\,=\,10$^{-4}$  we obtain $N({\rm H}_2)$\,=\,2.0\x10$^{22}$\,cm$^{-2}$ for the narrow PDR component.

Table~\ref{Line intensities Radex} collects the observed intensities and the corresponding values
calculated from the model. 
A source size of 120\arcsec, derived in Sect.~\ref{Section Observations}, is used to correct the antenna temperatures $T_{\rm A}^*$  
to 
Rayleigh-Jeans brightness temperatures $T_\mathrm{b}$ according to the  scaling in Eq.~\ref{solution}.
The estimated abundances and column densities are 
summarized in Table~\ref{radex abundances}. Note that the best-fitting model of the [C II]
$\lambda 157\,\mu$m line emission implies a density of carbon ions,
$n({\rm C}^+)$\,=\,7.7\,cm$^{-3}$. This is taken to be equal to the 
electron density, which implies an average electron fraction of
$X(e^-)$\,=\,1.9\x10$^{-5}$. This electron fraction is large enough that
electron-impact on polar molecules like H$_2$O must be competitive with
H$_2$ collisions in exciting rotational states. 
However, the results presented for NH$_3$ so far neglect electron collisions.

The model reproduces nicely all main beam temperatures except the ammonia   inversion transitions which are predicted to be 
lower than the observed intensities.
The \mbox{\13co(2\,--\,1)} transition is  also  predicted to be slightly lower than what is observed. These discrepancies may be caused by a co-existing, extended lower excitation gas component, which is not contributing much to the emission of the \13co(5\,--\,4) or  \mbox{\nh3(1$_0$\,--\,0$_{\,0}$)} lines. Even so,
the column densities of C, \13co and CS  agree well with   \citet{1994ApJ...425L..49P} and  \citet{1984ApJ...276..625S}, and the C abundance is also in agreement with \citet{2008ApJ...674.1015F}. 
The water abundance was previously estimated by 
\citet{2000ApJ.539.101.Snell.etal}  to  be 9\x10$^{-9}$, and by \citet{2000ApJ.539.119.Ashby} to be 2\x10$^{-8}$, both very similar to the abundance calculated by {\tt RADEX}.

The {\tt RADEX} model is also tested for higher transitions of \h2o and shows that some lines are sensitive to
various effects. For example, the intensity ratio of the 1661 and 1670\,GHz 
lines appears to be sensitive to the total column density.
When the internal radiation field is switched off in the models, the
intensities of the transitions involving the lowest rotational levels
increase slightly while the intensities of lines involving more highly
excited states are diminished: in particular, the 1661\,GHz line is
enhanced by radiative excitation.  

A series of excitation calculations with {\tt RADEX} at small optical depths
indicate that the intensity of the $2_{2,1}$\,--\,2$_{1,2}$ line at 1661\,GHz
is very sensitive to the total column density: evidently photon-trapping
is ineffective at column densities around \mbox{10$^{14}$\,\cmsq}~or
less, so that this line would appear in absorption if the opacity
in the 557\,GHz transition is small. 

The excitation of NH$_3$ also shows some interesting effects in these models.
Due to the rather
high density of C$^+$ ions, the reaction
of C$^+$ with NH$_3$ will destroy ammonia at a rate exceeding 
$10^{-8}$\,s$^{-1}$, which is rapid enough to affect the excitation of 
the metastable levels involved in several of the inversion transitions
near 23\,GHz. Indeed, the calculations indicate that the $(J,K)$\,=\,(3,3)
inversion transition is a weak maser \citep[cf.][]{1983A&A...122..164W}. In addition, NH$_3$ has strong
vibrational transitions in the mid- and near-infrared. When these are
included, the intensity of the 572\,GHz transition is somewhat suppressed
while the lowest inversion transitions are somewhat enhanced. 

 \begin{table}
  \caption{Results of the homogeneous {\tt RADEX} model for the narrow PDR component: column densities and abundances.$^{ a}$}
   \label{radex abundances}
    \centering
     \begin{tabular}{l c c  }

     \hline\hline
     \noalign{\smallskip}
Species & Column Density & Abundance  \\
 x & $N({\rm x})$ & $X({\rm x})$   \\
 & [cm$^{-2}$] &  \\
     \noalign{\smallskip}
     \hline
     \noalign{\smallskip}

 H$_2$O & $2.0\times 10^{14}$ & $ 9.8\times10^{-9}$  \\  
 H$_2^{18}$O & $<4.9\times 10^{12}$ & $<2.5 \times 10^{-10}$  \\ 
 NH$_3$ & $1.5\times 10^{14}$ & $7.5\times10^{-9}$   \\ 
 $^{12}$CO & $2.0\times 10^{18}$ & $ 1.0\times10^{-4}$   \\
 $^{13}$CO & $2.9\times 10^{16}$ & $ 1.4\times10^{-6}$  \\ 
 CS & $1.5\times 10^{14}$ & $ 7.5\times 10^{-9}$   \\
 C & $2.0\times 10^{17}$ & $ 1.0\times 10^{-5}$  \\
 C$^+$ & $9.0\times 10^{17}$ & 4.5$ \times 10^{-5}$   \\
 O & $6.3\times 10^{17}$ & $ 3.2\times 10^{-5}$   \\

     \noalign{\smallskip}
     \hline
     \end{tabular}
\begin{list}{}{}
\item[$^{{a}}$] All results for $T_{\rm K}$\,=\,55\,K, $n({\rm H}_2)$\,=\,4\x10$^5$\,cm$^{-3}$, $n(e^-)$\,=\,7.7\,cm$^{-3}$, $N({\rm H}_2)$\,=\,2.0\x10$^{22}$\,cm$^{-2}$.
\end{list}
   \end{table}

\begin{table} [!h]
\caption{Clumpy PDR model parameters.}
\label{PDR model parameters} 
\begin{tabular} {  c  l llllllll}
\hline
\hline
   \noalign{\smallskip}
size	&	$n_\mathrm{c}$(H)	&$n_\mathrm{ic}$(H)		&	$F^a$	 	&	$l_\mathrm{c}^b$ &	  
$T_\mathrm{gas}$ & $T_\mathrm{dust}$      \\
\,[pc] &[\cmcub]	&	[\cmcub]	& [\%]	 	 & [pc]	 	 &[K] &[K]  \\
   \noalign{\smallskip}
\hline
   \noalign{\smallskip}
0.5 & 2\x10$^6$	&	1\x10$^4$	&	8	 	&	0.03	 &9\,--\,40$^c$&18\--\,32$^c$\\ 
&&&&  &35\,--\,195$^d$&\\
   \noalign{\smallskip}
\hline
\end{tabular}
\begin{list}{}{}
\item$^{{a}}$Volume filling fraction of clumps. 
\item$^{{b}}$Clump size. 
\item$^{{c}}$Temperatures in clumps. 
\item$^{{d}}$Temperatures in interclump medium.
\end{list}
\end{table}

\section{Model results: clumpy PDR model} \label{subsection Spaans model}

A self-consistent physi-chemical 3D inhomogeneous model is used to compute the   temperature of the gas and dust, and the abundance distributions inside the clumpy cloud
\citep[detailed descriptions are found in][]{1996A&A.307.271.Spaans,1997A&A.323.953.Spaans.Dishoeck, 2005A&A.440.559.Poelman.Spaans}.
This clumpy PDR model  has three free parameters; the volume filling factor $F$ (fraction of the cloud that is occupied by clumps), a clump size $l_\mathrm{c}$, and  the clump-interclump ratio  $n_\mathrm{c}$(H)/$n_\mathrm{ic}$(H). The total hydrogen density $n_\mathrm{H}$\,=\,$n(\mathrm{H})$\,+\,2$n(\mathrm{H_2})$. In the dense clumps, hydrogen is entirely molecular, thus $n(\mathrm{H})$\,=\,2$n(\mathrm{H_2})$.
The   PDR model parameters used in this paper are given in  Table~\ref{PDR model parameters}.
We use a velocity dispersion of 1.2\,\kms ~\citep{1994ApJ...428..219Zhou} equivalent to a FWHM line width of 2.0\,\kms~for an optically thin line. With an opacity of about 5\,--\,10 the line is broadened by a factor of $\sim$1.5\,--\,2 to 3\,--\,4\,\kms, consistent with our observations.
The incident radiation field is taken to be $I_\mathrm{UV}$\,=\,140 with respect to the \citet{1978ApJS...36..595D} field.
The clumps are randomly distributed in the interclump medium
and the volume filling fraction is taken to be 8\% \citep{1997A&A.323.953.Spaans.Dishoeck, 2000ApJ.539.119.Ashby, 2005A&A.440.559.Poelman.Spaans}.
The grid size is 81x81x81, corresponding to a resolution of 0.006\,pc and a total size of 0.5\,pc (about 110\arcsec~at a  distance of 910\,pc). 

In the PDR computation, each clump is joined smoothly to the inter-clump
medium through a power-law $\propto$1/R$^2$ density distribution. This envelope contains
very little molecular material and the bulk of the impinging radiation
field is always absorbed by the clumps and not the inter-clump medium.
The definition of the mean H$_2$ density and the mean molecular abundances $X$ of species $x$ is
thus made through $<n>$\,=\, $Fn_\mathrm{c}$\,+\,$(1-F)\,n_\mathrm{ic}$ and $X_\mathrm{mean}$($x$)\,=\,$F\,X_\mathrm{c}(x)$\,+\,$(1-F)\,X_\mathrm{ic}(x)$~to allow an unambiguous
comparison with molecular line observations .

A high molecular hydrogen density of $\ga$1\x10$^6$\,\cmcub~ in the clumps is required
to match the intensities  observed by Odin and SWAS. Note that this clump density   is considerably higher than that used by \citet{2005A&A.440.559.Poelman.Spaans, 2006A&A.453.615.Poelman.Spaans}. 
The high density and clumpy medium suggested by our PDR model is, however,
also supported by previous density estimates. For example,  density variations between 10$^4$\,--\,10$^6$\,\cmcub~are   necessary to explain the apparent coincidence of the  \ion{C}{I}  and CO ridges as well as the distance between the  \ion{C}{I} ridge and the bright rim \citep{1992PASJ...44..391Hayashi}. 
Emission 
from numerous, small, dense \mbox{(10$^5$\,--\,2\x10$^6$\,\cmcub)} and optically thick clumps of gas,
where the number density of clumps decreases with distance from the CS cloud centre close to IRS,
 is the best explanation of the CS column density variations, as well as the observed opacities and temperatures obtained from multitransition observations of CS 
 \citep{1984ApJ...276..625S, 1994ApJ...428..219Zhou}.
As  molecular hydrogen density in the interclump medium we use 5\x10$^3$\,\cmcub.

\begin{table} [!h]
\caption{Resulting abundances with respect to $n_\mathrm{H_2}$ for the narrow PDR components of water and ammonia from the clumpy PDR model.
}
\label{abundances result from PDR model} 
\centering
\begin{tabular} { ll l l }
\hline
\hline
   \noalign{\smallskip}
Species & Comp. &$n_\mathrm{H_2}$ & Abundance    \\
&	& [\cmcub]	 &        \\
   \noalign{\smallskip}
\hline
   \noalign{\smallskip}
\h2o	& Interclump &	5.0\x10$^3$	&8.9\x10$^{-10}$	\\  
	& Clump&	1.0\x10$^6$&	4.8\x10$^{-8}$ \\  
	& Mean 	& 8.5\x10$^4$  & 4.6\x10$^{-9}$ \\ 
   \noalign{\smallskip}
\nh3	& Interclump &	5.0\x10$^3$	&1.1\x10$^{-10}$	\\    
	& Clump&	1.0\x10$^6$&	4.5\x10$^{-8}$ \\  
	& Mean 	&  8.5\x10$^4$& 3.7\x10$^{-9}$\\  

   \noalign{\smallskip}
\hline
\end{tabular}
\end{table}

\begin{figure}[h] 
\resizebox{\hsize}{!}{\includegraphics[scale=0.15]{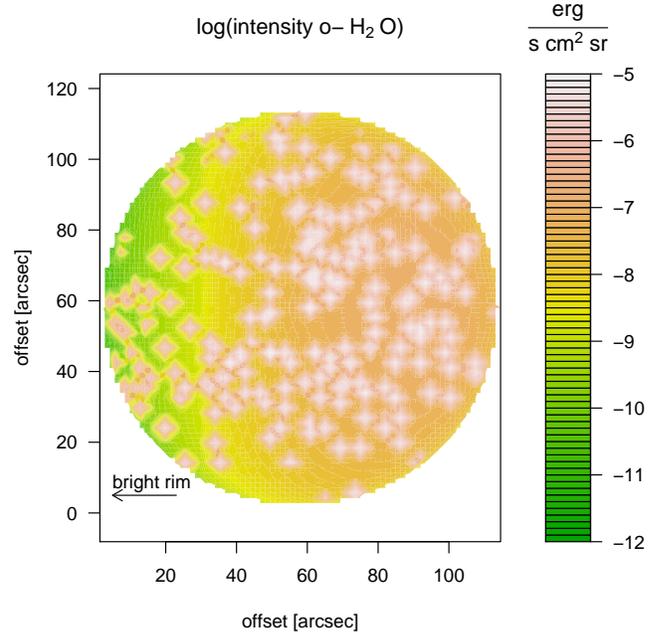}} 
\caption{Result of the PDR high-density model. The  water intensity
is shown in a logarithmic scale.  The ionizing star is located on the left side of the displayed region.}
\label{spaans contour intensity highdens model}
\end{figure}

The resulting abundances, with respect to H$_2$, in the clumps and the inter-clump medium, as well as  a mean value over the whole region, are found in Table~\ref{abundances result from PDR model}.  The mean value is about half than found by {\tt RADEX}. In the clumps, 
the water   abundances vary between   10$^{-7}$\,--\,10$^{-9}$  with an  average of 4.8\x10$^{-8}$.
The low  density  in the interclump medium results in low water
abundances   between  1\x10$^{-11}$\,--\,5\x10$^{-9}$ with an average of 8.9\x10$^{-10}$.  
Ammonia is found to have about the same average abundance as water, 4.3\x10$^{-8}$, in the clumps,  while lower by almost an order of magnitude in the  inter-clump medium, 1.1\x10$^{-10}$.

\begin{table} [!h]
\caption{Observed and predicted  $T_\mathrm{A}^*$ in the PDR for  \mbox{\emph{o}-\h2o(\trans)}    with SWAS and Odin  at the central position.
}
\label{spaans Ta} 
\centering
\begin{tabular} {  l ll}
\hline
\hline
   \noalign{\smallskip}
 Satellite        & Obs.  $T_\mathrm{A}^*$& Pred. $T_\mathrm{A}^*$ \\
 &                       [mK]     &              [mK]       \\
   \noalign{\smallskip}
\hline
   \noalign{\smallskip}
 SWAS   &186& 120\\
 Odin &416&345 \\
   \noalign{\smallskip}
\hline
\end{tabular}
\end{table}

The result from our PDR model is   used as input to a three-dimensional multi-zone escape probability method which computes the level populations of  \emph{ortho}- and \emph{para}-\h2o  (up to $\sim$350\,K), CO and \nh3,  as well as line intensities and opacities \citep{2005A&A.440.559.Poelman.Spaans, 2006A&A.453.615.Poelman.Spaans}. The excitation due to dust emission is fully included.
The dust temperature in the model varies between 18 and 32\,K. However,  due to geometry the highest $T_\mathrm{dust}$
is only achieved at the edge of the PDR and the dust component fills only a small part of the region. The bulk of the dust has a temperature of about 25\,K and the dust continuum opacity $\lambda$100\,$\mu$m is 0.02 or smaller. 
Results from this code differ  at most by 10\%
compared to other Monte Carlo/Accelerated Lambda Iteration
 codes.
Fig.~\ref{spaans contour intensity highdens model} shows the predicted water intensities on a logarithmic scale.
Note that the ionizing star is located on the left side of the displayed region.
The peak emission is found about \mbox{70\,--\,80\arcsec}~from the bright rim,   in agreement with our observations of all species.

The resulting model intensities are then used to calculate the
expected antenna temperature via
\begin{eqnarray}\label{TA}
\lefteqn{T_\mathrm{A}^* = \frac{1}{2}\,\frac{SA_{eff}}{k\,\Delta\nu}=\frac{1}{2}\,\frac{b\,\eta_\mathrm{bf}\,\Omega_\mathrm{mb}}{k\,\Delta \nu}\,\frac{\lambda^2 }{ \Omega_\mathrm{A}}=  {} }
\nonumber\\
&&{}
 =  \frac{1}{2}\,\frac{b\,\lambda^2}{k\,\Delta\nu}\, \eta_\mathrm{bf}\,\eta_\mathrm{mb} 
= \frac{b\,\lambda^3}{2\,k\,\Delta\upsilon}\,\eta_\mathrm{mb}\,\eta_\mathrm{bf}
\,\, \,[\mathrm{K}],
\end{eqnarray}

\begin{figure}[h] 
\resizebox{\hsize}{!}{\includegraphics[scale=0.5]{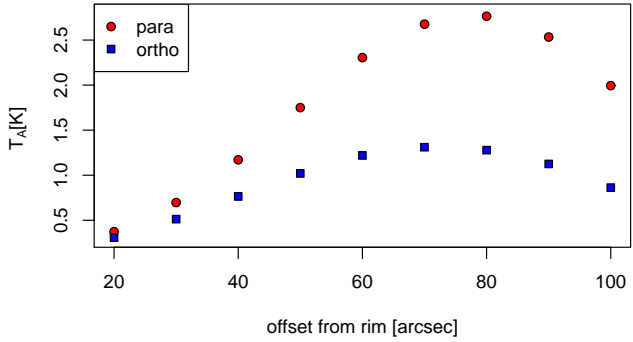}} 
\caption{Predicted   antenna temperatures of the  \emph{ortho}-\h2o(\trans) and \emph{para}-\h2o(\mbox{1$_{1,1}$\,--\,0$_{0,0}$}) transitions  for the forthcoming Herschel-HIFI observations  along a cut 20\,--\,100\arcsec~from the bright rim.}
\label{HIFI predictions from spaans model}
\end{figure}

where  $S$ is the integrated flux density in [erg s$^{-1}$\,\cmsq], the  intensity $b$ is in [erg\,s$^{-1}$\,\cmsq\,sr$^{-1}$], $A_\mathrm{eff}$ is the effective antenna area  $\eta_\mathrm{A}\,A_\mathrm{geo}$\,=\,$\lambda^2/\Omega_\mathrm{A}$ [cm$^2$],
$\Delta \nu$ is the observed line width  in Hz, $\Omega_\mathrm{mb}$ is the main beam solid angle in [sr], 
$\Delta \upsilon$ is the observed line width  in cm\,s$^{-1}$, and the  wavelength $\lambda$ is in cm.
We 
convolve the intensities to the Odin beam, and in addition predict the intensities of the forthcoming observations of  the Herschel Space Observatory using a telescope diameter of  350\,cm and a beam-efficiency of  0.7.
The observed and modelled SWAS and Odin antenna temperatures for the \mbox{\trans}~transition  are provided in Table~\ref{spaans Ta} for the central position. In addition,  the predicted  antenna temperatures  for Herschel-HIFI  of both the \emph{ortho}- and \emph{para}-\h2o  \trans~and \mbox{1$_{1,1}$\,--\,0$_{0,0}$} transitions along a cut from  the bright rim are shown in Fig.~\ref{HIFI predictions from spaans model}.

The model \emph{ortho}-water opacities are relatively low with a unweighted mean opacity of 17 over the entire source,  with a range between 10$^{-5}$\,--\,800. 

The antenna temperatures from the PDR model  agree well with our observations, especially considering uncertainties in source and clump size, as well as the volume filling fraction, which is the main uncertainty and can be varied to obtain slightly higher or lower $T_\mathrm{A}^*$.  
The  dust background continuum is also  lower than in the {\tt RADEX} model, which
diminishes the effect of radiative excitation of water.
A clumpy medium with high clump densities above 10$^6$\,\cmcub~is, however, necessary to produce model antenna temperatures close to the observed values. This is also supported by the resulting low density of 3\x10$^4$\,\cmcub~obtained by the dust model in Sect.~\ref{subsection RADEX} assuming a homogeneous, spherical cloud, which is about 7\% of the density in our best fitting model.

\section{Summary and conclusions}  \label{section conclusions}

We have used the Odin satellite to observe water, ammonia and carbon-monoxide in the well-known molecular cloud S140.
We have simultaneously observed   five-point strip maps across the  bright rim in S140  of the \emph{ortho}-\h2o \trans~and  the \emph{ortho}-\nh3 1$_0$\,--\,0$_{\,0}$ transitions, as well as \mbox{CO(5\,--\,4)} and \mbox{\13co(5\,--\,4)}. The  \nh3 transition has never previously been observed in S140. Observations of \water18 in the central position resulted in a non-detection at a rms level of 8\,mK.
As support observation we also mapped \mbox{\13co(1\,--\,0)}    with the Onsala 20m telescope.

Like CO, the water line-profile  is dominated by emission from a NW\,--\,SE  outflow, however, mainly in the red wing. 
Strong self-absorption is seen in the optically thick CO emission, while no obvious signs are seen in the \emph{ortho}-\h2o, \emph{ortho}-\nh3 or the almost optically thin \13co line profiles.
In  addition to the outflow, our water line shows emission from a more extended NE\,--\,SW elongated PDR.
Both these components originate approximately around our central position.   No additional emission closer to the bright rim or further into the molecular cloud is detected.
Close to the bright rim the   temperature is most likely too high for a detection of our transition with   an upper state energy of 61\,K. Instead, higher-lying transitions, observable with the Herschel Space Observatory, will have their peak intensity shifted towards  the bright rim \citep{2005A&A.440.559.Poelman.Spaans, 2006A&A.453.615.Poelman.Spaans}.  Even closer to the bright rim at a few magnitudes of $A_\mathrm{V}$, water is, however, photo-dissociated by the UV field.

The \emph{ortho}-\nh3 emission seems to emanate from the same high density clumps in the PDR and the outflow as water, but  also shows  additional emission further into the cloud where the ambient gas temperature drops    to  about 30\,K.  
Compared to water  in the central position, ammonia has a weaker outflow emission in the red wing although similar in the blue outflow. 
Close to the bright rim, where the outflow contribution to the emission is very low,  the water and ammonia line profiles are almost identical, suggesting an origin in the same gas and velocity fields of the PDR. The \13co line also shows a very similar line profile as \h2o and \nh3 in this position with a narrow line width of $\sim$3\,\kms. 

Abundances, with respect to  H$_2$, in the PDR  are estimated  both with an enhanced version of the homogeneous {\tt RADEX} code and with a clumpy PDR model.
This   model points to   low mean water and ammonia abundances  in the PDR of 5\x10$^{-9}$ and 4\x10$^{-9}$, respectively. 
In the high-density clumps both the average water and ammonia abundances increase to 5\x10$^{-8}$.
To match the observed PDR antenna temperatures  with Odin and SWAS, a clumpy medium is required by the model with a high  
molecular hydrogen density in the clumps of $\ga$1\x10$^6$\,\cmcub.
The resulting {\tt RADEX} mean abundances are twice as high: 1.0\x10$^{-8}$  and 8\x10$^{-8}$ for water and ammonia, respectively, using a molecular hydrogen density of 4\x10$^{5}$\,\cmcub~and a kinetic temperature of 55\,K.
The differences most likely arise from the uncertainty in density, beam-filling, and volume filling of the clumps. 
The opacity    of the narrow PDR component of the \trans~transition is constrained by the narrow line width, and is  estimated  by  {\tt RADEX} to be $\sim$7.
The PDR  model also confirm  a  low water opacity with an unweighted mean opacity of 17   and a model range of $\sim$10$^{-5}$\,--\,800.
The mean outflow water abundance,  derived from a simple two-level approximation, is higher than in the PDR by at least one  order of magnitude,
$\sim$2\x10$^{-8}$\,--\,2\x10$^{-7}$.

Predictions of antenna temperatures for observations with   HIFI  are given by our PDR model 
of the \emph{ortho}- and \emph{para}-\h2o \trans~and 1$_{1,1}$\,--\,0$_{0,0}$ transitions 
for  nine positions across the bright rim, and are found to peak around 70\,--\,80\arcsec~from the dissociation front in agreement with our observations.

\begin{acknowledgements}
We thank   Per Bergman, Magnus Gustavsson,  Matthijs Klomp and also Steve Shore and the organizers  of the   A\&A and EDP sciences school: Scientific Writing for Young Astronomers for helpful discussions.
Generous financial support from the Research Councils and Space Agencies in Sweden, Canada, Finland and France is gratefully acknowledged.

\end{acknowledgements}

\bibliographystyle{aa-package/bibtex/aa}
\bibliography{references}

\Online

\appendix
\section{{\tt RADEX} -- construction of a dust model} \label{on-line dust model construction}

In order to relate the molecular column densities, 
$N({\rm x})$ of species x, to fractional abundances, $X({\rm x})$\,=\,$N({\rm x})/N({\rm H}_2)$,  a uniform, homogeneous sphere of diameter \mbox{$L$\,=\,$N({\rm H}_2)/n({\rm H}_2)$} is assumed here. The
adopted physical diameter of the PDR corresponding to an  
angular diameter of
120\arcsec~(Sect.~\ref{Section results}) at a distance of 910\,pc is 0.53\,pc. This is assumed to be
equal to the line-of-sight depth.

The observed intensity of the 
continuum is used to estimate the internal radiation field sensed by
the molecules. We construct a simple model of the broad-band 
spectrum at submm and far-infrared wavelengths in order both to 
characterize the internal radiation and to estimate the total 
column densities of dust and   hydrogen.
\citet{1983ApJ...271..625S} measured the far-infrared emission of S140
and found a peak flux density of the order of $10^4$\,Jy slightly
shortward of $\lambda$100\,$\mu$m  in a 49\arcsec~beam. \citet{1995A&A...298..894M.Minchin.etal}
presented total broad-band fluxes in a $1\farcm 5\times 1\farcm 5$ 
box. We represent the latter results with a 
two-component model of thermal emission by dust over a solid angle
of $\Omega = 1.9\times 10^{-7}$\,sr. The main component has a dust 
temperature $T_{\rm dust}$\,=\,40\,K and a long-wavelength ($\lambda > 40
\,\mu$m) form of the opacity law 
$$ \tau_{\rm dust} = 0.0679 (100/\lambda)^{1.2} $$
where $\lambda$ is the wavelength in $\mu$m. The opacity law is smoothly
matched to a standard interstellar extinction law at shorter wavelengths,
which is also used to describe the second component at $T_{\rm dust}$\,=\,140\,K. 
The opacity of the first dust component corresponds to a
visual extinction $A_V$\,=\,58.8\,mag. The second component has a smaller
optical depth $A_V$\,=\,0.023\,mag, but is assumed to cover the same solid
angle. 
In addition, the mid-infrared measurements of \citet{Ney}  have been adapted in order to specify the radiation field at 
even shorter wavelengths.
In the calculations, the molecules are assumed to be exposed
to an average intensity of continuous radiation 
$$  I_{\nu} = B_{\nu}(T_{\rm CMB}) + \eta {{f_{\nu}^{\, \rm dust}}\over
{\Omega}} $$
where $B_{\nu}$ is the Planck function, $T_{\rm CMB}$\,=\,2.73\,K is the 
temperature of the cosmic background radiation, 
$f_{\nu}^{\, \rm dust}$ is the flux
density of the 2-component dust model, 
$\Omega$\,=\,1.9\x10$^{-7}$\,sr, and $\eta$\,=\,0.72
is a dilution factor to scale the brightness of the dust source to the
larger beam area of the Odin measurements. It is important to keep in
mind that we observe this strong far-infrared radiation; therefore, the 
co-extensive molecules must sense it also.

For the adopted interstellar extinction law and a standard gas/extinction ratio, 
$$ 2 N({\rm H}_2) = 1.6\times 10^{21} A_V \,\,{\rm cm}^{-2}, $$
the adopted dust model implies $N({\rm H}_2)$\,=\,4.7\x10$^{22}$\,\cmsq~ and an average density 
$n({\rm H}_2)$\,=\,2.9\x10$^4$\,cm$^{-3}$ over the source
size $L$\,=\,0.53\,pc. 
This average density is, however, inconsistent
with the observed molecular line emission in large beams 
\mbox{($\theta \geq 1'$).} 
Although a uniform {\tt RADEX} model can be constructed based
upon this density, 
the  line-center optical depths
of the pure rotational lines of H$_2$O
and NH$_3$ would be of the order of 200 and
100, respectively. Such large opacities would imply significant 
line broadening through saturation of the emission, which conflicts with the
observed narrow profiles of $\sim$3\,\kms. \\ \\ \\ \\ \\ \\ \\

\section{Figures and Tables}

\begin{figure}[h] 
\centering
\includegraphics[scale=0.45]{0930fig8.eps}
\caption{Odin observations of   \water18 in the central position.  
} 
 \label{spectrum h2o18}
\end{figure}

\begin{figure}[h] 
\centering
\includegraphics[scale=0.45]{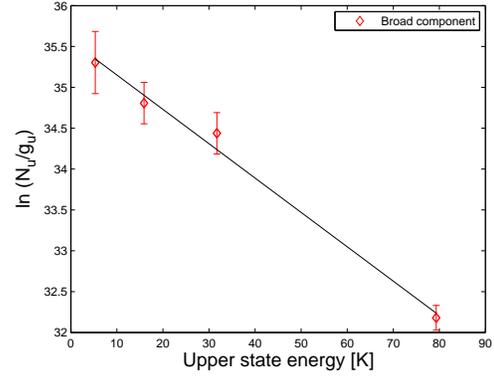}
\caption{Rotation diagram of  the broad component  of $^{13}$CO(1\,--\,0) with the Onsala 20-m telescope, $J$\,=\,2\,--\,1 and  $J$\,=\,3\,--\,2 from  \citet{1993A&A...277..595M.Minchin.etal}, and  $J$\,=\,5\,--\,4 with Odin,
producing $T_\mathrm{ROT}$\,=\,24$\pm$2\,K and $N_\mathrm{ROT}$\,=\,(2.5$\pm$0.4)\x10$^{16}$\,\cmsq.  
} 
 \label{Rotdiagram 13co outflow}
\end{figure}

\begin{figure}[h] 
\centering
\includegraphics[scale=0.45]{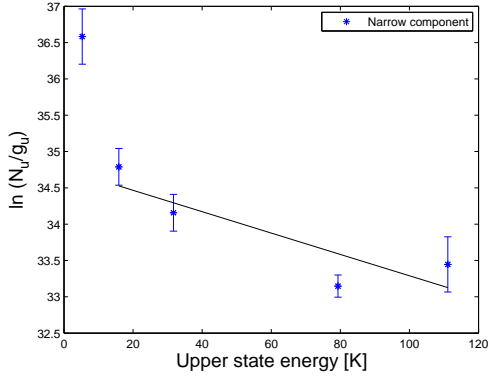}
 \caption{
Rotation diagram of  the narrow component  of \13co(2\,--\,1) and  $J$\,=\,3\,--\,2 from \citet{1993A&A...277..595M.Minchin.etal},  $J$\,=\,5\,--\,4 
with Odin, and   
$J$\,=\,6\,--\,5  from \citet{1993ApJ...405..249Graf},    producing
$T_\mathrm{ROT}$\,=\,69$\pm$27\,K and $N_\mathrm{ROT}$\,=\,(3.2$\pm$1.8)\x10$^{16}$\,\cmsq. 
$^{13}$CO(1\,--\,0)  is    not included in the fit.
} 
 \label{Rotdiagram 13co 4 lines PDR}
\end{figure}

\begin{table*}
\caption{Observed transitions and their parameters$^{{a}}$ in S140 with the Odin satellite in a five point NE-SW strip.}
\label{Odin table}
\begin{tabular} { l l l l l l l r r r  rl}
\hline
\hline
Species		& Transition 	 & Freq  	& $E_\mathrm{u}$	&$A$-coeff & $\Delta \upsilon$&	Pos. & $\upsilon_{\mathrm{LSR}}$ & $T_\mathrm{peak}$
& $\int$ $T^{\,*}_{\mathrm{A}}$ d$\upsilon$       &rms   &note  \\
	                      &	          & [GHz]	 &	[K] 	& [s$^{-1}$] &[kms$^{-1}$] &&[kms$^{-1}$]& [mK] & [K kms$^{-1}$]&[mK]  &\\
\hline

 H$_2$O	& 1$_{1,0}$\,--\,1$_{0,1}$& 556.936&61.0&3.46e-3 &0.27   
 
      &  1      &-8.0 & 207 &  0.64&39&  \\ 
      
&&&&&&2& -6.9 & 517 & 2.74&68&\\ 

&&&&&&3& -6.6 & 632 &3.34 &19&\\ 

&&&&&&4& -7.0 & 379 &1.84 &57&\\ 

&&&&&&5&  --  &--   &   -- & 44& No detection. \\

H$_2^{18}$O 	& 1$_{1,0}$\,--\,1$_{0,1}$& 	547.676	&60.5	&3.29e-3&0.34&3&-- &-- &-- &$\sim$8 &	 No detection.\\ 

CO 	& 5\,--\,4& 	576.258		&83.0 &1.22e-5&0.32&1&-5.4 &6\,680& 47.1&67 &	\\

&&&&&&2& -5.0 &11\,760  &94.2 &47&\\
&&&&&&3& -5.5 & 13\,240 &108.8 &43&\\
&&&&&&4& -6.6 & 8\,690 & 60.1 &50&\\
&&&&&&5& -6.9 & 2\,220 & 10.1 &160&\\

$^{13}$CO 	& 5\,--\,4& 	550.926	&79.3	&1.10e-5&0.27&1&  -7.8 & 2\,740& 8.29 &115& \\  

&&&&&&2& -7.4 & 5\,930 & 21.4&81&\\
&&&&&&3& -7.4 & 7\,110 & 27.7&61&\\
&&&&&&4& -7.6 & 3\,640 & 13.2&119&\\
&&&&&&5& -8.5 & 1\,190  &2.5&291&\\ 

NH$_3$ 	& 1$_{0}$\,--\,0$_{0}$& 	572.498		&27.5& 1.61e-3&0.32&1&
                       -8.0 &  444  & 1.18&99&	 \\	
		       
&&&&&&2& -7.4  & 467 & 1.83 &33&\\
&&&&&&3& -7.4 & 588 & 2.66 &29&\\
&&&&&&4& -7.3  & 381  & 1.86 &36&\\
&&&&&&5& --  & --  &--  &91&No detection\\
\hline
\end{tabular}
\begin{list}{}{}
\item$^{{a}}$ Transition\,=\,the quantum numbers for the transition; Freq\,=\,rest frequency of the transition;   $E_\mathrm{u}$\,=\,the upper state energy; $A$coeff\,=\,the Einstein $A$-coefficient; $\Delta \upsilon$\,=\,the velocity resolution; Pos\,=\, the strip position from NE to SW. $\upsilon_\mathrm{LSR}$\,=\,the peak LSR velocity;  $T_\mathrm{peak}$\,=\,the observed peak temperature of the transition;   $\int$ $T^{\,*}_{\mathrm{A}}$ d$\upsilon$ \,=\,the integrated intensity from the observed spectra not corrected for beam-efficiency or beam-filling; rms\,=\,noise.
\end{list}
\end{table*}

\begin{table*} [!h]
\caption{$^ {13}$CO Gaussian fits$^{a}$.  $T_\mathrm{b}$ uses a source size for the PDR (narrow component) of 120\arcsec $\rightarrow \eta_\mathrm{bf}$\,=\,2, and  a source size for the broad outflow component of 85\arcsec $\rightarrow \eta_\mathrm{bf}$\,=\,3.} 
\label{result_tableCO} 
\centering
\begin{tabular} {  l r r r r r r r}
\hline
\hline
Pos.  	&	$\upsilon_{\mathrm{LSR}}$& err  & $T_\mathrm{A^*}$ & $T_\mathrm{b}$ & err&  	$\Delta \upsilon$ & err  \\
	 &	  [km\,s$^{-1}$]&[km\,s$^{-1}$] &[mK]&[mK]&[mK]&[km\,s$^{-1}$]	&[km\,s$^{-1}$] \\
\hline

1&-7.7&0.04&2\,665&5\,920	&68&2.8&0.08\\

2&-7.3&0.02&5\,877&	13\,060	&	56&3.3&0.04\\

3&-7.3&0.01&6\,610&	14\,690	&199&3.2&0.07\\
3&-6.8&0.33&612&  2\,040		&207&8.2&1.54\\

4&-7.3&0.03&3\,620&		8\,040	&66&3.5&0.07\\
5&-8.2&0.25&1\,124&		2\,500	&282&2.1&0.62\\

\hline
\end{tabular}
\begin{list}{}{}
\item[$^{a}$] Gaussian fits (including errors) to the spectra in 5 positions. For most positions two Gaussians are needed to fit the spectra. Parameters: $\upsilon_\mathrm{LSR}$\,=\,the LSR velocity at the peak temperature, amp\,=\,the amplitude of the Gaussian fit, $\Delta \upsilon$\,=\,FWHM line width.
\end{list}
\end{table*}

\begin{table*} [!h]
\caption{H$_2$O Gaussian fits$^{a}$. $T_\mathrm{b}$(PDR) uses a source size of 120\arcsec  $\rightarrow \eta_\mathrm{bf}$\,=\,2,
while $T_\mathrm{b}$(outflow) uses a source size of 85\arcsec  $\rightarrow \eta_\mathrm{bf}$\,=\,3.} 
\label{result_tableh2o} 
\centering
\begin{tabular} {  l r  r r r r r r  }
\hline
\hline
Pos.  	&	$\upsilon_{\mathrm{LSR}}$& err  & $T_\mathrm{A^*}$ & $T_\mathrm{b}$ &err&  	$\Delta \upsilon$ & err    \\
&	 	  [km\,s$^{-1}$]&[km\,s$^{-1}$] &[mK]&[mK]&[mK]&[km\,s$^{-1}$]	&[km\,s$^{-1}$]\\

\hline
 1& -7.5  &0.10& 180& 400&12& 3.1&0.2 \\

 2&-6.9& 0.08&340&756&28&3.5& 0.3  \\   
 
  &-5.4&0.39&171&570&27&8.8&0.7 \\

 3&-7.1&0.06&416&924&30&3.1&0.20 \\ 
 
  &-6.1&0.24&213&710&31&8.8& 0.60\\

 4&-7.1&0.08&316&702& 44 &3.5&0.4  \\       
 
 &-6.6&0.8&78&260 &45 &10.0& 4.1 \\

\hline
\end{tabular}
\begin{list}{}{}
\item[$^{a}$] Notation as in Table~\ref{result_tableCO} .
\end{list}
\end{table*}

\begin{table*} [!h]
\caption{NH$_3$ Gaussian fits$^{a}$. $T_\mathrm{b}$(PDR) uses a source size of 120\arcsec $\rightarrow \eta_\mathrm{bf}$\,=\,2, 
while $T_\mathrm{b}$(outflow) uses a source size of 85\arcsec  $\rightarrow \eta_\mathrm{bf}$\,=\,3.} 
\label{result_tablenh3} 
\centering
\begin{tabular} {  l r  r r r r r r }
\hline
\hline
Position  	&	$\upsilon_{\mathrm{LSR}}$& err  & $T_\mathrm{A^*}$ & $T_\mathrm{b}$ & err&  	$\Delta \upsilon$ & err   \\
&	 	  [km\,s$^{-1}$]&[km\,s$^{-1}$] &[mK]&[mK]&[mK]&[km\,s$^{-1}$]	&[km\,s$^{-1}$]\\

\hline
 1& -7.5  &0.14& 438 &	973 &	46& 2.6&0.3 \\   
 
 2&-7.9& 0.25&412& 916	&200&3.3& 0.9 \\  
 
&-6.1& 3.4&111&	370 &160&5.8& 2.6 \\ 
 
 3&-7.6&0.12    &487   &1\,080 	&	90   &3.3&0.50 \\ 
 
&-6.4&1.4   &100  & 333	&	92  &8.5&4.3 \\  

 4&-7.4&0.23  &   321& 713	&	185 &   3.9   & 1.2 \\
 
&-6.8   &2.6     &    61   &  203	&	192    &   8.7     & 12.1 \\

\hline
\end{tabular}
\begin{list}{}{}
\item[$^{a}$] Notation as in Table~\ref{result_tableCO} .
\end{list}
\end{table*}

\begin{figure}[h] 
\centering
 \resizebox{\hsize}{!}{\includegraphics{0930fg11.eps}}
\caption{Gaussian fits to  \13co(5\,--\,4) at the central  position. The widths, amplitudes and centre velocities are 3.2\,\kms~and 8.2\,\kms;   6.610\,K and 0.612\,K; $-$7.3\,\kms~and $-$6.8\,\kms, respectively.
} 
 \label{13co 5-4 center 2 Gauss}
\end{figure}

\begin{figure}[h] 
\centering
 \resizebox{\hsize}{!}{\includegraphics{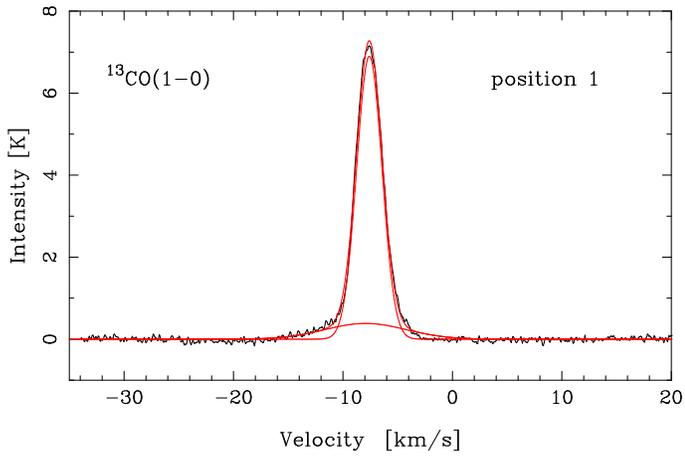}}
\caption{Gaussian fits to the convolved (to the Odin 126\arcsec~beam) spectra of \13co(1\,--\,0) at  position 1. The widths, amplitudes and centre velocities are 2.7\,\kms~and 8.6\,\kms;   6.982\,K and 0.395\,K; $-$7.6\,\kms~and $-$8.0\,\kms, respectively.
} 
 \label{13co 1-0 2 Gauss}
\end{figure}

\begin{figure}[h] 
\centering
 \resizebox{\hsize}{!}{\includegraphics{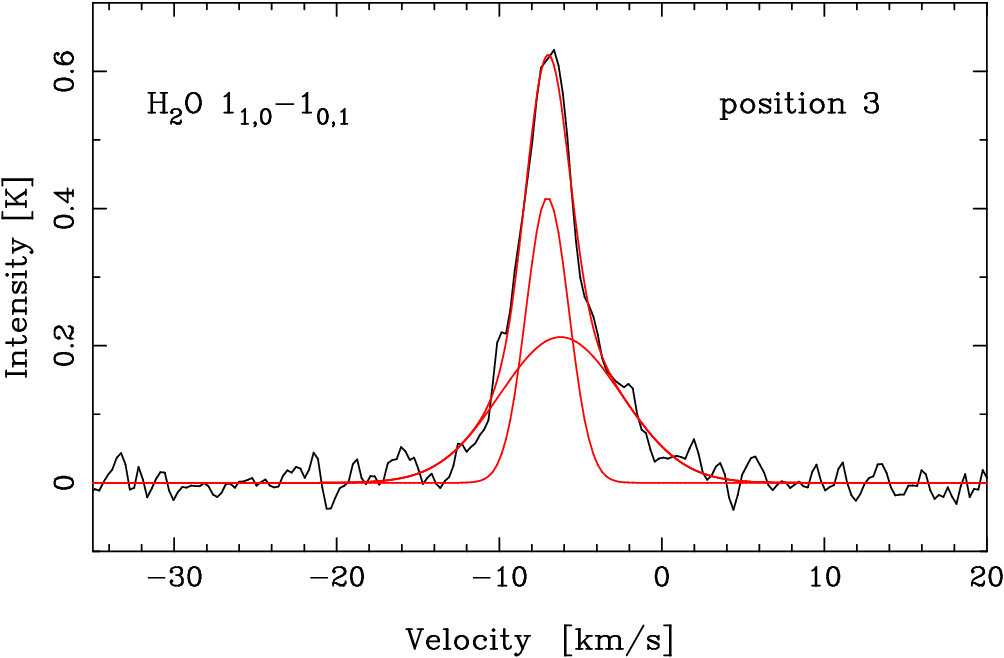}}
\caption{Gaussian fits to \h2o at the central position. The widths, amplitudes and centre velocities are 3.1\,\kms~and 8.8\,\kms;   416\,mK and 213\,mK; $-$7.1\,\kms~and $-$6.1\,\kms, respectively.
} 
 \label{h2o center 2 Gauss}
\end{figure}

\begin{figure}[h] 
\centering
 \resizebox{\hsize}{!}{\includegraphics{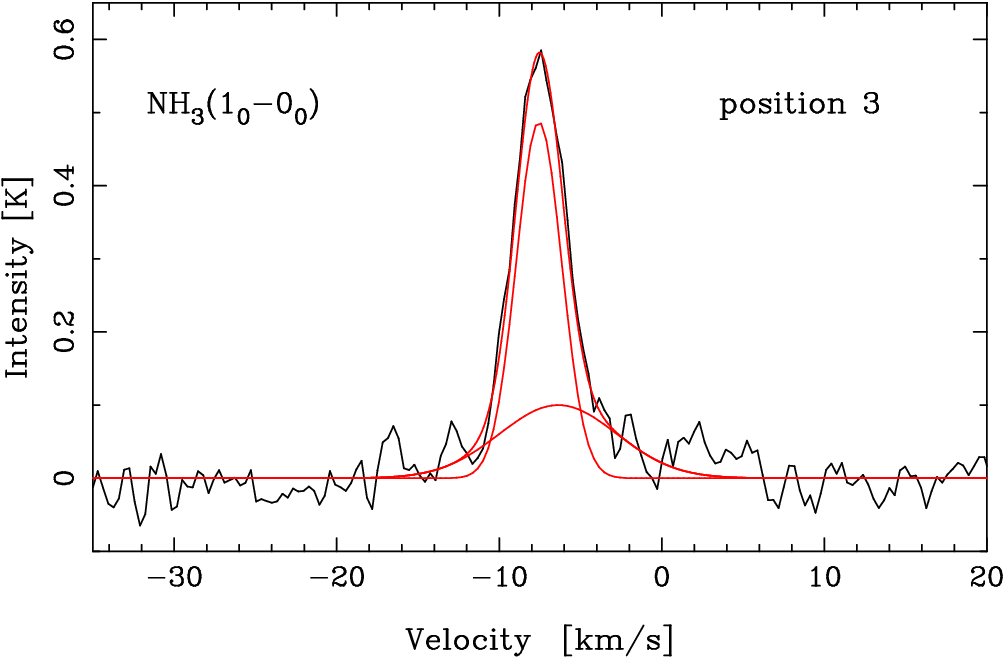}}
\caption{Gaussian fits to \nh3 at the central position. The widths, amplitudes and centre velocities are 3.3 \kms~and 8.5 \kms;   487~mK and 100~mK; -7.6~\kms~and -6.4~\kms, respectively.
} 
 \label{nh3 center 2 Gauss}
\end{figure}

\end{document}